\documentclass[useAMS,usenatbib]{mn2e}
\usepackage{natbib}
\usepackage{graphicx}
\newcommand{\msun}{\mbox{$\,{\rm M}_\odot$}}

\title[Tidal Tails of Star Clusters]{Tidal Tails of Star Clusters}

\author[A.H.W. K\"upper, P. Kroupa, H. Baumgardt and  D.C. Heggie]{Andreas
  H.W. K\"upper$^{1}$\thanks{E-mail: \mbox{akuepper@astro.uni-bonn.de} (AHWK);
    \mbox{pavel@astro.uni-bonn.de} (PK); \mbox{holger@astro.uni-bonn.de} (HB); \mbox{d.c.heggie@ed.ac.uk} (DCH)}, Pavel Kroupa$^1$, Holger Baumgardt$^1$ and Douglas C. Heggie$^{2}$\\
$^{1}$Argelander Institut f\"ur Astronomie (AIfA), Auf dem H\"ugel 71, 53121 Bonn, Germany\\
$^{2}$University of Edinburgh, School of Mathematics and Maxwell
Institute for Mathematical Sciences, King's
Buildings, Edinburgh EH9 3JZ, UK}

\begin{document}

\date{Accepted \ldots. Received \ldots; in original form \ldots}

\pagerange{\pageref{firstpage}--\pageref{lastpage}} \pubyear{2009}

\maketitle

\label{firstpage}

\maketitle

\begin{abstract}
Based on recent findings of a formation mechanism of substructure in
tidal tails by \citet{Kuepper08a} we investigate a more comprehensive set of $N$-body models of star clusters on orbits about a Milky-Way-like potential. We find that the predicted epicyclic overdensities arise in any tidal tail no matter which orbit the cluster follows as long as the cluster lives long enough for the overdensities to build up.

The distance of the overdensities along the tidal tail from the cluster centre depends for circular orbits only on the mass of the cluster and the strength of the tidal field, and therefore decreases monotonically with time, while for eccentric orbits the orbital motion influences the distance, causing a periodic compression and stretching of the tails and making the distance oscillate with time. We provide an approximation for estimating the distance of the overdensities in this case.

We describe an additional type of overdensity which arises in extended tidal tails of clusters on eccentric orbits, when the acceleration of the tidal field on the stellar stream is no longer homogeneous. Moreover, we conclude that a pericentre passage or a disk shock is not the direct origin of an overdensity within a tidal tail. Escape due to such tidal perturbations does not take place immediately after the perturbation but is rather delayed and spread over the orbit of the cluster. All observable overdensities are therefore of the mentioned two types. In particular, we note that substructured tidal tails do not imply the existence of dark-matter sub-structures in the haloes of galaxies.
\end{abstract}

\begin{keywords}
galaxies: kinematics and dynamics -- galaxies: star clusters -- methods: analytical -- methods: $N$-body simulations -- galaxies: haloes -- cosmology: dark matter\end{keywords}

\section{Introduction}
Since an increasing number of tidal structures are being discovered around the Milky Way (e.g. \citealt{Odenkirchen01, Belokurov06, Grillmair06b, Grillmair08, Juric08, Keller09, Koch09}), there is a profound interest for a deeper physical understanding of tidal tails. Tidal tails offer direct constraints on the ongoing dissolution process of star clusters as well as on the last few Gyr of their evolution. Furthermore, their alignment, shape and structure give a unique opportunity to reconstruct the cluster's orbit in unprecedented detail and thus probe our understanding of the Milky-Way potential.

From the simple assumption that a star cluster has a nearly constant mass-loss rate on its orbit about the galactic centre, its tidal tails are often expected to have a more or less homogenous and smooth structure. Substructure in the tidal tails is usually only expected in time-variable tidal fields when the mass-loss rate varies with time.

However, in a recent paper (K\"upper, Macleod \& Heggie 2008; hereafter KMH) we analytically predicted and numerically proved the formation of substructure in tidal tails even within a constant tidal field. This effect is due to an epicyclic motion of stars evaporating from the cluster instead of the more common approximation of linear motion along the tails (for details see KMH). Like a standing wave this motion leads to statistical overdensities if it is performed by a stream of stars.

We were able to show that these \emph{epicyclic overdensities} arise at a distance of many tidal radii along the orbit from the cluster centre and also gave a simple relation between this distance and the tidal radius of the corresponding cluster. The numerical verification was done for a fairly simple $N$-body model, using a point-mass galaxy and a star cluster without a stellar-mass spectrum, to avoid all disturbing influences.

More recently, this effect has also been studied by \citet{Just08} in a more general context, showing that, even in a Milky-Way-like potential and with a cluster consisting of stars drawn from a mass function, these over- and underdensities arise and can give valuable information on the current state of the cluster.

In KMH we found that the prerequisite for the formation of epicyclic overdensities is that a majority of escaping stars leaves the cluster with velocities slightly above the escape velocity through one of the two Lagrange points. If the scatter in escape conditions is too large, the statistical overdensities are expected to vanish. For constant tidal fields this condition has been shown to be well fulfilled, but for time-variable tidal fields this has not been investigated yet.

For star clusters on eccentric orbits or on orbits involving periodic disk shocks the tidal field varies with time and the internal evolution of the cluster is perturbed. Thus, the scatter in escape conditions may be increased such that epicyclic overdensities may not be able to arise. How large the influence of the perturbation is depends on the orbital parameters. Therefore, we here investigate a comprehensive set of $N$-body simulations of globular clusters on different orbits about a Milky-Way-like potential, and study the formation of substructure within the tidal tails (Sec.~\ref{sec:ts}). Within the models we also look for overdensities in the tails which can be assigned to a tidal variation and an associated change in the mass-loss rate. But first a few theoretical considerations have to be set out to explain the various aspects of escape and, in this context, explain our methodology (Sec.~\ref{sec:Theory}).

\section{Theory of escape}\label{sec:Theory}
Understanding  the formation of tidal tails, and especially of substructure within those tails, is relatively easy in the ideal case of a star cluster on a circular orbit about a spherically symmetric galaxy as studied in KMH. Escape from such a cluster can be theoretically described in the framework of epicyclic theory in which the equations of motion of escaping stars are fairly simple. KMH showed that stars, when they evaporate with small velocities from the cluster through one of the Lagrange points, move on  oscillatory orbits along the tidal tails. These oscillations lead to periodic stellar over- and underdensities when they are performed by a constant stream of escapers. But how does this change for more complex clusters in time-variable tidal fields?

Since the formation of substructure in the tails, as described in KMH, is due to stars evaporating from the cluster, and in a constant tidal field evaporation is mass-loss driven by two-body relaxation \citep{Kuepper08b}, we argue that the more stars escape with \textit{evaporative escape conditions} from the cluster, i.e. escape with low speed through one of the Lagrange points, the more pronounced are the overdensities and the easier it is to observe overdensities beyond the first one.

In time-variable tidal fields this does not necessarily mean that a) escape has to be due to two-body relaxation, since pericentre passages during eccentric orbits and disk shocks also create escapers, as they enhance the mass-loss rate in general \citep{Vesperini97, Baumgardt03} or that  b) escape has to happen from the Lagrange points, as the distance of the Lagrange points from the cluster centre may vary more quickly on eccentric orbits than the cluster profile does.

On the other hand, time variations in the tidal field, and associated variations in the mass-loss rate, are also the most frequently suggested formation mechanism for substructure in tidal tails \citep{Dehnen04}. However, this mechanism has not been convincingly proven in numerical experiments yet. Therefore, we need to study the mass-loss rate and the escape conditions of the investigated models in detail to understand the origin of detected substructure.

\subsection{Mass-loss rate}
\subsubsection{Constant tidal fields}
The time scale of evaporation in constant tidal fields is the relaxation time, $t_{rh}$, as the mass loss of a tidally limited star cluster can be approximated by
\begin{equation}\label{eq:dMdt}
 \frac{dM}{dt} \propto  t_{rh}^{-3/4}
\end{equation}
\citep{Baumgardt01}. The median two-body relaxation time is furthermore given by \citep{Spitzer87}
\begin{equation}\label{eq:trel}
 t_{rh} = 0.138 \frac{N^{1/2}R_h^{3/2}}{G^{1/2}m^{1/2}\ln{\Lambda}},
\end{equation}
where $N$ is the number of stars, $R_h$ is the half-mass radius, $G$ the gravitational constant, $m$ the mean mass ($M/N$) and $\ln{\Lambda}$ the Coulomb logarithm where $\Lambda$ is of order $0.11 N$ in the case of equal masses \citep{Giersz94}.

For clusters on circular orbits the tidal field is static and the cluster will dissolve as a result of two-body relaxation and the interactions of stars with the tidal field, where the mass-loss rate is determined approximately by the ratio of half-mass radius to tidal radius \citep{Gieles08}. For such time-independent tidal fields the tidal radius scales with $M^{1/3}$, hence decreases rather slowly with ongoing mass loss. Since the change in the half-mass radius also happens rather slowly, the mass-loss rate is approximately constant during one revolution about the galactic centre. Hence, substructure in the tidal tails can only arise through the epicyclic motion of escaping stars.

\subsubsection{Time-variable tidal fields}
If the cluster moves on an eccentric orbit or if the cluster periodically crosses the galactic disk then additional energy is put into the cluster on an orbital time scale and the mass-loss rate is expected to vary during one revolution.

As the galactocentric distance of the cluster periodically changes on an eccentric orbit, the tidal radius is not monotonically decreasing any more but varies during one period. Furthermore, because the structure of the cluster may not be able to change rapidly enough, a part of the cluster may (temporarily) lie beyond the tidal radius when the cluster is at pericentre,  where the tidal radius reaches its minimum. This induces a strong pull on those outlying stars, increasing on average the stellar energies. When the tidal radius increases again, a fraction of these once outlying stars may be recaptured but now with increased energy. The rest will be lost from the cluster. This whole process is often loosely referred to as bulge shocking (for a detailed discussion of tidal shocking see for example \citealt{Gnedin99}).

During a disk shock the cluster is temporarily compressed on a time scale smaller than the internal dynamical time scale. As a consequence of this, all stars get a kick, accelerating some stars (mostly the stars in the outer parts) to velocities above the escape velocity, and these are then likely to leave the cluster. This behaviour can be described, for example, in the formalism of the first- and second-order theory of disk shocks as reviewed in \citet{Gnedin99}. Here, the mean energy changes, $\langle\Delta E\rangle$ and $\langle\Delta E^2\rangle$, of the stars within the cluster can be expressed as
\begin{eqnarray}
 \langle \Delta E \rangle & \propto & g^2_mr^2,\label{eq:dE}\\
 \langle \Delta E^2 \rangle & \propto & g^2_mv^2r^2,\label{eq:dE2}
\end{eqnarray}
where $g_m$ is the maximum vertical gravitational acceleration of the disk while $r$ and $v$ are the positions and velocities of the corresponding stars. The dependence on $r^2$ makes clear that, on the one hand a cluster is more affected by a shock when it is more extended, and on the other hand that the outer stars are most likely to be pushed to energies above the escape energy.
Moreover, the energy shifts from equations \ref{eq:dE} and \ref{eq:dE2} become less influential with increasing galactocentric radius of the disk shock since the vertical gravitational acceleration decreases. \citet{Vesperini97} found that disk shocks at solar radius and beyond are almost negligible for realistic globular cluster configurations.

\subsubsection{Delayed escape}\label{ssec:delay}
In both types of time-dependent tidal-field, pericentre passages and disk shocks, the stars most affected  are at radii comparable to the (apogalactic) tidal radius. At these radii the orbital time scale of stars within the cluster is on the order of the orbital time scale of the cluster within the galaxy.

For constant tidal fields \citet{Fukushige00} have shown that a star which is pushed to an energy just slightly above the escape energy cannot escape from the cluster easily but has to pass near one of the two Lagrange points,  as the potential barrier is lowest there. This significantly increases the escape time of stars from a cluster \citep{Baumgardt01}.

What happens in the case of time-dependent tidal fields  has not been investigated yet. Nevertheless, for somewhat non-circular orbits it is likely that,  for many affected stars,  escape will be delayed until well after the pericentre passage or the disk shock. This delay in escape of a specific star may even last several revolutions,  since the star may need several orbits within the cluster to escape, as in the circular case.
Hence the time variation of the tidal field need not lead to detectable overdensities within the tidal tails as suggested by \citet{Dehnen04} if the fraction of delayed escapers is large and escape events are spread in time. However, the formation of epicyclic overdensites will still be possible, even if a majority of stars leaves the cluster as a result of tidal variations, as long as the escape conditions of the escaping stars are evaporation-like, i.e. escape with low velocities.

\subsection{Escape conditions}
\subsubsection{Constant tidal fields}
\citet{Giersz94b}, \citet{Baumgardt02} and \citet{Kuepper08b} were able to show that for clusters (in constant tidal fields) the escape velocities of the majority of escaping stars follow a log-normal distribution when normalized to the specific velocity dispersion of the cluster.

This behaviour can be understood in terms of two-body relaxation: the velocity distribution within the cluster is almost Maxwellian and stars in the high-velocity tail of this distribution are deemed to leave the cluster as their velocities exceed the escape velocity of the cluster. The latter is furthermore linked to the velocity dispersion for clusters in virial equilibrium. Hence, the ratio of the escape velocity of an escaping star to the velocity dispersion of the cluster at that time is expected to be constant. And since the velocity dispersion changes rather slowly with time, stars in the tidal tails have quite similar escape conditions (which we call here evaporative escape conditions) leading to thin and dynamically cold tidal tails with observable epicyclic overdensities.

\subsubsection{Time-variable tidal fields}
As mentioned above, tidal variations affect the velocities of cluster stars. Hence, for strong tidal variations the velocity distribution does not necessarily have to be nearly Maxwellian any more. Furthermore, the tidal radius and thus the escape velocity of the cluster changes during one revolution about the galactic centre. In addition, the velocity dispersion periodically varies with time as the cluster gets compressed or oscillates. Finally, the cluster can even be temporarily out of virial equilibrium.

From these  considerations we may expect that the escape conditions have a larger spread around a mean value and may even deviate from log-normal and become asymmetric. In the extreme case of a cluster which gets completely disrupted through a single shock the scatter in escape conditions will be maximal. Thus, broad and dynamically hot tidal tails without epicyclic overdensities are expected from this crude picture.
The intermediate regime, from perfectly evaporative escape conditions to maximally perturbed conditions, will be the subject of our parameter-space study.

\subsection{Number- and orbital-velocity distribution}\label{sec:Method}
\begin{figure}
  \includegraphics[width=84mm]{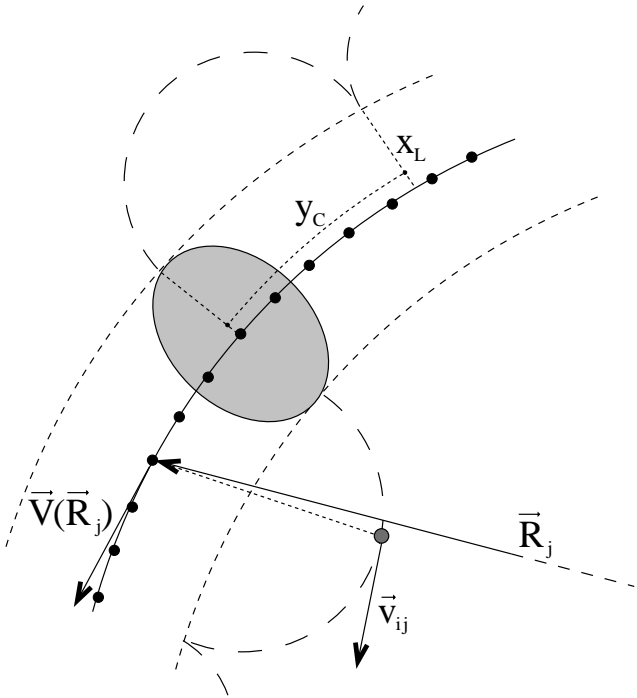}
  \caption{Sketch of the analysis setup. The grey ellipse represents the cluster while the small grey circle shows a star in the tidal tails with index $i$ on an epicyclic motion within the tails. The solid line with black knots gives the orbit of the cluster, where the knots show equidistant points along this orbit. $\vec{R}_j$ is the galactocentric radius of the knot with index $j$, and $\vec{V}\left(\vec{R}_j\right)$ the orbital velocity of the cluster at this position respectively. As star $i$ is closest to the knot at  $\vec{R}_j$ it is given the index $ij$ and counted in bin $j$. Its velocity is $\vec{v}_{ij}$. $x_L$ is the tidal radius, and $y_C$ gives the length of the epicyclic loops which are indicated by the long-dashed lines.}
  \label{velocities}
\end{figure}
We are going to analyse the tidal tails of our models in terms of number- and orbital-velocity distributions, in order to try to find the origin of any observed overdensities and relate them to the tidal radius of the cluster.  In other words, motivated by the results of KMH, we investigate the number distribution of stars along the tidal tails and compare it to the average orbital velocities of stars at that position within the tails. Epicyclic motion of a constant stream of stars leaving the cluster creates over- and underdensities at periodic distance intervals. For constant tidal fields the distance of the density enhancements only changes when the conditions of escape are modified, e.g. when the tidal radius changes. Furthermore, the oscillatory (epicyclic) movement also creates a signal in the orbital velocity of the stars along the tails, as was first discovered by \citet{Capuzzo05} in numerical simulations: places where the stars move fastest (slowest)  should correspond to the positions of underdensity (overdensity) (cf. KMH).

To analyse the clusters in a consistent way, all stars are binned along the orbit in bins of size 25 pc. For this purpose, we calculate points, $\vec{R}_j$, along the orbit of the cluster at separations of 25 pc and assign the tail stars to the point which they are closest to. The resulting distribution gives the number density of stars as a function of distance from the cluster centre along the orbit. In addition, the average orbital velocity with respect to the cluster within the $j$th bin, $v_j$, is measured as
\begin{equation}\label{eq:v}
v_j = \left(\frac{1}{N_j}\sum^{N_j}_{i=0}v_{ij}\right)-V\left(\vec{R}_j\right),
\end{equation}
where $v_{ij}$ is the  speed of the $i$th star in the $j$th bin, $N_j$ the number of stars in the $j$th bin and $V\left(\vec{R}_j\right)$ is the speed the cluster will have or had at the position $\vec{R}_j$ of the orbit (Fig.~\ref{velocities}). Furthermore, the velocity dispersion in each bin is calculated with
\begin{equation}\label{eq:sigma}
\sigma_j = \sqrt{\overline{v_j^2}-\overline{v_j}^2},
\end{equation}
where $\overline{v_j}$ is the mean velocity in the $j$th bin and $\overline{v^2_j}$ the mean squared velocity respectively.

\subsection{Tidal radius and distance $y_C$}\label{sec:rt}
In KMH we theoretically predicted and numerically verified a relation between   the positions of the overdensities, $y_C$, and the tidal radius of the cluster, $x_L$. For a cluster in a point-mass galactic potential the position of the first overdensity is
\begin{equation}\label{eq:yc_pm}
y_C = \pm 12\pi x_{L},
\end{equation}
but for the general case the relation gets more complicated. The distance of the first overdensity is then given by
\begin{equation}\label{eq:yc}
y_C = \pm \frac{4\pi \Omega}{\kappa}\left( 1-\frac{4\Omega^2}{\kappa^2}\right) x_L,
\end{equation}
where $\Omega$ is the angular velocity of the cluster on its orbit about the galactic centre and $\kappa$ is the so-called epicyclic frequency (see KMH). The latter comes from epicycle theory, where the motion of a star is approximated by the circular motion of a guiding centre about the galactic centre and an oscillation of the star about this guiding centre (cf. \citealt{Binney87}).

In this framework the epicyclic frequency for a cluster moving in a planar orbit is given by
\begin{equation}\label{eq:kappa}
 \kappa^2 = \frac{\partial^2\Phi}{\partial R^2}+3\Omega^2.
\end{equation}
Here, $\Phi$ is the galactic potential and $R$ is the galactocentric radius of the cluster. The epicyclic frequency gives a quantification of the steepness of the galactic potential in the radial direction and the strength of the centrifugal force, and therefore naturally appears in the equation for the tidal radius (as the distance of the Lagrange points from the cluster centre is usually called):
\begin{equation}\label{eq:rtide}
 x_L^3 = \frac{GM}{4\Omega^2-\kappa^2},
\end{equation}
where $G$ is the gravitational constant. The epicyclic frequency obviously has a large influence on the value of $x_L$. While the relation $\kappa = \Omega$ holds for point-mass galaxies, the value for a Milky-Way-like potential may vary between approximately $0.5 \Omega$ and $2 \Omega$. Fortunately, the flat rotation curve of the Milky Way implies that there is a wide range of galactocentric distances within the disk where $\kappa$ is about
\begin{equation}\label{eq:kappamw}
\kappa \simeq 1.4\Omega,
\end{equation}
(cf. Fig.~3 in \citealt{Just08}).

As mentioned above, this theory only holds for nearly circular orbits. In the case of arbitrary orbits in complicated potentials these approximations break down and no simple solutions to the equations of motion can be found.

In the following analysis the tidal radius is always determined by combining equation \ref{eq:kappa} and \ref{eq:rtide}, i.e.
\begin{equation}
 x_L^3 = \frac{GM}{\Omega^2-\partial^2\Phi/\partial R^2}.
\end{equation}
The angular velocity is furthermore calculated using
\begin{equation}
\Omega =  \frac{|\vec{R}\times\vec{V}|}{R^2},
\end{equation}
where $\vec{R}$ is the position of the cluster with respect to the galactic centre and $\vec{V}$ its velocity.

The basic quantities which we investigate (bound mass, mean stellar mass, etc.) will be calculated for all stars within this theoretical radius, and the distance of the detected overdensities relative to the density centre compared to it. This distance, $y_C$, should only be related to the tidal radius if the overdensities are of epicyclic origin.

\subsection{$y_C$ in the case of time-dependent tidal fields}\label{ssec:yc of t}
There is no theory for the distance of the epicyclic overdensities, $y_C$, in the case of time-dependent tidal fields yet. But we can estimate the distance between the cluster and the epicyclic overdensities, even when the system gets periodically accelerated and decelerated on an eccentric orbit or through a galactic disk, if we make further assumptions:

\begin{enumerate}
\item Firstly, we have to consider the fact that the acceleration of the cluster and of the stars in the tails will change with time. Consequently, the distance between any two points moving along the orbit is not constant but changes according to the orbital phase. What stays constant due to angular momentum conservation though, is the difference between the times at which the two points pass a certain point of the orbit. We can therefore convert the distance $y_C$ into a conserved time difference $\Delta t$. For constant tidal fields $y_C$ can be expressed as
    \begin{equation}\label{eq:dt}
        y_C = V \Delta t,
    \end{equation}
where $V$ is the magnitude of velocity of the cluster on its circular orbit about the galaxy and $\Delta t$ is the time difference at which the cluster and the overdensity pass a certain point of the orbit.

\item For eccentric orbits or orbits including disk shocks the cluster and its tails get periodically stretched and compressed as the velocity of the cluster and the tails changes during a period. The distance between two points which are separated by a given amount of time, $\Delta t$, can be written in a Taylor expansion as
    \begin{equation}\label{eq:dt2}
        y_C(t) = V(t) \Delta t + \frac{1}{2}A(t)\Delta t^2 + O(\Delta t^3),
    \end{equation}
where $A$ is the acceleration at the given point. Since $V$ and $A$ are well known for our models, the separation of two points which would be at a constant distance in a constant tidal field, e.g. the cluster and the first-order overdensity, can be calculated for a time-variable tidal field.

\item Furthermore, we shall assume that the time variations of the tidal field are sufficiently fast  that the cluster cannot adapt to the changing environment, but rather behaves as if it experiences a single mean tidal field along its orbit. For this purpose we shall average the galactocentric distance, $R$, and the orbital velocity, $V$, and use these quantities to calculate an average $y_C$ using eq.~\ref{eq:yc} and assuming a circular orbit at the given mean galactocentric distance.

The results of these assumptions will be checked in Sec.~\ref{ssec:fgpeo}, but the following argument explains why assumption (iii) is reasonable. The length of the epicycle, which escaping stars follow after they leave the cluster vicinity, depends on the offset of the star from the cluster centre perpendicular to the orbit (see KMH or \citealt{Just08} for a more detailed discussion). In a constant tidal field this offset is the tidal radius. But for time-variable tidal fields, the stars do not necessarily have to escape from the tidal radius, as this radius might change faster than the cluster can adapt to the changing tidal conditions. In fact, stars escape from what we might refer to as the ``edge'' of the cluster, whose radius changes little during one orbit. Indeed, recent studies show that stars escape preferentially from a radius approximately equal to the apogalactic tidal radius of the cluster; furthermore they have low velocities, making the formation of epicyclic overdensities  likely, even in the case of non-circular orbits (K\"upper, Kroupa, Baumgardt \& Heggie, in preparation).

\item Finally, by applying eq.~\ref{eq:dt} we can estimate a time by which the cluster and the overdensity should be separated on their orbit. With this approximate value of $\Delta t$ we have all the ingredients to predict the behaviour of $y_C$ with time during one orbital period using eq. \ref{eq:dt2}.
\end{enumerate}

\subsection{Mean drift velocity}
The time escaping stars need to complete one epicycle, e.g. the time that passes from the time a star leaves the cluster until it reaches the first overdensity, is given by
\begin{equation}
 t_C = 2\pi/\kappa,
\end{equation}
and in KMH we also showed that the distance between the overdensities is given by equation \ref{eq:yc}. Thus, we can determine a mean drift velocity of the stars within the tidal tails with respect to their guiding centre by using
\begin{equation}\label{eq:vc}
v_C = \frac{y_C}{t_C}=\pm 2\Omega \left(1-\frac{4\Omega^2}{\kappa^2}\right)x_L,
\end{equation}
where the minus (plus) sign holds for the leading (trailing) tail. Hence, the mean drift velocity depends on the tidal-field properties through $\kappa$ and $x_L$. This equation is only valid for circular orbits, though. For circular orbits in the galactic disk of the Milky Way the mean drift rate is about
\begin{equation}
v_C \simeq \pm 2\Omega x_L=\pm\left(4GM\Omega\right)^{1/3},
\end{equation}
where we used equations \ref{eq:rtide} and \ref{eq:kappamw}. For a constant mass-loss rate the density within the tails is therefore increasing with decreasing mass as the angular velocity of the cluster is constant whereas the tidal radius decreases monotonically. Hence, the stars propagate more slowly along the tails the less massive the cluster is.

In time-dependent tidal fields the mean tidal conditions used in the formalism of Sec.~\ref{ssec:yc of t} may be applied to get a rough estimate of the mean drift velocity. But due to the compression and stretching of the tails throughout the orbit a drift velocity in a literal sense is not meaningful.

The mean drift velocity, $v_C$, which is calculated in the rotating reference frame of the leading or trailing tail, can be converted into the average orbital velocity of Sec.~\ref{sec:Method}, which is measured in the reference frame of the cluster, by making a correction, $\widetilde{V}$, for the difference in orbital velocity of the guiding centre of the corresponding tail to the orbital velocity of the cluster. This correction emanates from the left part of the right-hand side of eq.~4 in KMH, which gives the offset of the guiding centre of the tails from the cluster centre, multiplied by the angular velocity $\Omega$, and reads
\begin{equation}
\widetilde{V} = \pm \Omega\frac{4\Omega^2}{\kappa^2}x_L,
\end{equation}
where the plus (minus) sign holds for the leading (trailing) tail. The average orbital velocity in the tails with respect to the cluster, $\overline{v}$, is then given by
\begin{eqnarray}\label{eq:vccorr}
\overline{v} &=& v_C + \widetilde{V} = \pm 2\Omega \left(1-\frac{4\Omega^2}{\kappa^2}\right)x_L \pm \Omega\frac{4\Omega^2}{\kappa^2}x_L\\
 &=& \pm\Omega\left( \frac{4\Omega^2}{\kappa^2}-2\right) x_L.
\end{eqnarray}
This equation gives the mean value of the velocity profile along the tidal tails which will be measured in the following investigation using eq.~\ref{eq:v}.

\section{Numerical investigation}\label{sec:ts}
\begin{table}
\begin{minipage}{84mm}
\centering
 \caption{Overview of all computed models. In the first column the section is given where the particular model is discussed. $R_{apo}$ gives the apocentre distance from the galactic centre and $R_{peri}$ the pericentre distance, $incl$ the inclination of the orbit with respect to the galactic disk and $\epsilon$ the eccentricity of the orbit (eq.~\ref{eq:ecc}). All clusters have initially 65536 stars and a mass of about 20000 $\msun$. The models with inclination of 90 deg to the disk always cross the disk at $R_{gal}^{peri}$:  the eccentricity of these orbits is simply due to the asymmetric galactic potential.
}
\label{table1}
\begin{tabular}{ccccc}
\hline
 Sec. & $R_{apo} [kpc]$ &  $R_{peri} [kpc]$ & $incl$ [deg]& $\epsilon$ \\
\hline
\ref{ssec:pmgco} & 8.50 & 8.50 & -\footnote[1]{Cluster is moving in a point-mass galactic potential.} & 0.00 \\
\hline
\ref{ssec:fgpco} & 8.50 & 8.50 & 0 & 0.00 \\
\ref{ssec:fgpeo} & & 5.10 & 0 & 0.25 \\
   & & 2.83 & 0 & 0.50\\
   & & 1.21& 0 & 0.75 \\
\ref{ssec:fgppo} & 9.27 & 8.50  &  90 & 0.04 \\
\hline
\ref{ssec:fgprgal} &  4.25 & 4.25 & 0 & 0.00\\
 &   & 2.55 & 0 & 0.25 \\
   & &1.42& 0 & 0.50 \\
   & &0.61 & 0 & 0.75 \\
 & 5.24 &4.25 &90 & 0.10\\
\hline
 & 12.75& 12.75 & 0 & 0.00\\
   & &7.65 & 0 & 0.25\\
   & &4.25& 0 & 0.50\\
 &  &1.82&  0 & 0.75\\
 &   13.27 & 12.75& 90 & 0.00\\
\hline
 & 17.00 &17.00& 0 & 0.00\\
 &   &10.20& 0 & 0.25 \\
 & & 5.67& 0 & 0.50\\
 & & 2.43& 0 & 0.75\\
 & 17.44&17.00 &  90 & 0.01\\

\end{tabular}
\end{minipage}
\end{table}
The aim of this section is to trace the behaviour of tail formation from the ideal case of KMH to the complex case of a realistic star cluster. Therefore, we start with a star cluster like the one of KMH and, as a first step, add a mass spectrum to the cluster stars (Sec. \ref{ssec:pmgco}) to see if it affects the formation of epicyclic overdensities. Thereafter, we change the point-mass galactic potential to a Milky-Way potential with bulge, disk and halo (Sec. \ref{ssec:fgpco}). This will then be our reference model for the rest of this paper, i.e. we shall compare each model with this one. Afterwards, we study the influence of different orbital types on the formation of substructure. First, the orbit is changed from circular to eccentric (Sec. \ref{ssec:fgpeo}) and then a set of computations is made with circular orbits but perpendicular to the disk (Sec. \ref{ssec:fgppo}) to get an idea of the importance of disk shocks on cluster evolution. Additionally, all these kinds of orbits are computed at galactocentric distances of 4250, 8500, 12750 and 17000 pc.  First the clusters at an initial galactocentric radius of 8.5 kpc are discussed, but the effect of changing this radius is studied separately in Sec. \ref{ssec:fgprgal}. An overview of the models can be found in Table~\ref{table1}.

\subsection{The setup}
The initial clusters are chosen such that they can represent the Milky-Way globular cluster Palomar~5 (which is itself about 10 Gyr old) during the last 3 Gyr of evolution, i.e. rather 'fluffy' with a ratio of half-mass radius to tidal radius of about 0.2 \citep{Harris96}. We do this because Pal~5 is the only globular cluster with prominent tidal tails in which significant substructure has been found \citep{Odenkirchen01, Odenkirchen02, Odenkirchen03}. Furthermore, there are hardly any clusters present in the Milky Way which are more vulnerable to tidal influences than Pal~5. Most globular clusters in the Milky Way are much more compact and therefore can resist pericentre passages or disk shocks much better. Hence, the complete set gives us an idea of the maximum importance of a wide range of realistic tidal influences on star clusters and their tidal tails.

All computations were performed with the collisional $N$-body code \textsc{NBODY4} \citep{Aarseth99, Aarseth03} on the GRAPE-6A supercomputers at the AIfA \citep{Fukushige05}. The internal initial conditions of the chosen clusters are all similar: the clusters contain 65536 stars drawn from a canonical, two-part power law initial mass function \citep{Kroupa01, Kroupa08} ranging from 0.1 to 1.2 $\msun$; this results in a total cluster mass of about 20000 $\msun$, which is comparable to the mass of Pal~5 3 Gyr ago as estimated by \citet{Dehnen04}. The stars follow a Plummer density profile with a half-mass radius chosen such that $R_h/R_{tide}=0.2$ for a circular orbit, i.e. all clusters at a given apogalactic distance use the same cluster setup to ensure comparability among each other.

The clusters are set up as isolated clusters and then put into the tidal field, such that there is a tide-dependent unbound fraction of stars at the very beginning, which is in most cases negligible and mostly results in a slightly higher mass-loss rate at the beginning of the computations.

The clusters do not contain any primordial binaries because binaries would significantly enhance the CPU demand without adding to the problem at hand. Furthermore, stellar evolution is neglected to focus on dynamical effects. After setup the calculations were carried out for 4 Gyr if possible.

Only one cluster was computed for a point-mass galactic potential; for all other clusters a Milky-Way potential consisting of bulge, disk and spherical halo as described in \citet{Allen91} was used. For the latter potential, the equation for $x_L$ (eq. \ref{eq:rtide}) cannot be used unrestrictedly, because it only holds for spherical symmetric potentials in which the total angular momentum is naturally conserved or for orbits lying in the plane of symmetry, i.e. for orbits in the disk, for which the component of the angular momentum perpendicular to the disk is conserved. Nevertheless, even for the orbits which do not lie in the plane of symmetry the equation is valid to a sufficient degree of accuracy because the variation of the total angular momentum is only of order 10\%.

\subsection{Point-mass galaxy, circular orbit}\label{ssec:pmgco}
\begin{figure}
\includegraphics[width=84mm]{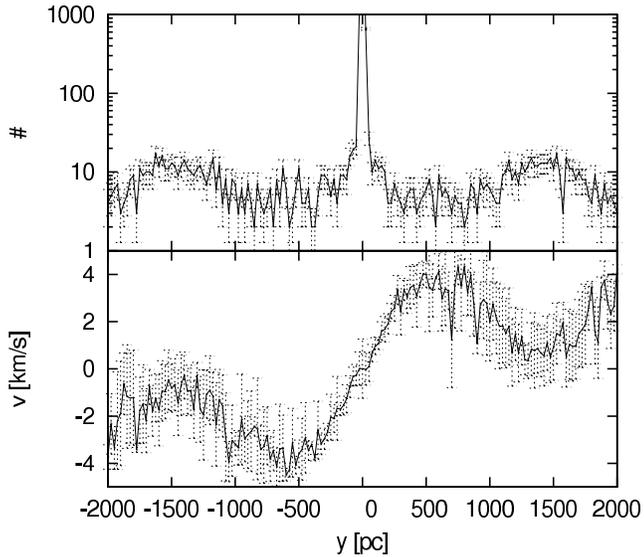}
  \caption{Number distribution of stars along the tidal tails in bins of 25 pc (upper panel) and corresponding mean velocities of the stars within the bins (with respect to the orbital velocity of the cluster, eq.~\ref{eq:v}) for the cluster in the point-mass galactic potential (discussed in Sec. \ref{ssec:pmgco}). The snapshot was taken when the cluster had a mass of about $16300 \msun$ at $t=2.0$ Gyr, which means a theoretical tidal radius of roughly $x_L = 33$ pc and a predicted distance of the first-order overdensity of 1250 pc. Error bars in the upper panel are Poisson errors and in the lower panel they show the velocity dispersion within the corresponding bin (eq.~\ref{eq:sigma}).}
  \label{y_pm}
\end{figure}
The step from a cluster consisting of single-mass stars to a multi-mass cluster is necessary, although no deviation from the theoretical predictions of KMH is expected, since the solutions to the equations of motion of stars leaving a cluster through its Lagrange points (equations 4-6 therein) do not depend on the masses of the stars. The only thing to check is whether the evaporative escape conditions are still fulfilled. But, as discussed in KMH, in constant tidal fields this condition is better fulfilled the more stars there are within the cluster, and here we are dealing with 64k stars compared to $N = 1000$ in KMH. Furthermore, \citet{Just08} already proved the formation of epicyclic over- and underdensities for this kind of cluster. Nevertheless, we perform this experiment to confirm our methodology.

Fig.~\ref{y_pm} shows a snapshot of the system at $t=2.0$ Gyr when the cluster has a bound mass of about $16300 \msun$, i.e. a tidal radius of 33 pc. The cluster clearly shows overdensities at the predicted positions of about $y_C=\pm 12\pi x_L \simeq 1250$ pc and even shows signs of the second- and third-order overdensities at $\pm 2y_C$ and $\pm 3y_C$ respectively (which are not visible in the plot because the plotted range of $y$ is restricted to facilitate comparison with similar plots for other clusters). Also the orbital velocity with respect to the cluster in the lower panel of Fig.~\ref{y_pm} clearly shows the predicted signal of periodic acceleration and deceleration. Moreover, the average orbital velocity in this figure is in good agreement with the predictions of eq.~\ref{eq:vccorr}, yielding an average orbital velocity of about $\overline{v} = \pm1.7$kms$^{-1}$ with respect to the cluster for the values of $\Omega$ and $x_L$ given in the caption of Fig.~\ref{y_pm}.

From this figure we can also see that most escaping stars have a velocity which is larger than the escape velocity, by the following argument. Eq.~\ref{eq:yc} gives the distance of the first epicyclic loop for a star which passes the Lagrange point with vanishing velocity; this distance is about 1250 pc in Fig.~\ref{y_pm}. If a star has an excess velocity, $v$, at the moment of escape then the length of the epicycle increases (see eq. 22 in \citealt{Just08}) as
\begin{equation}\label{eq:yc+v}
 y_C = \pm \frac{4\pi \Omega}{\kappa}\left( 1-\frac{4\Omega^2}{\kappa^2}\right) \left(x_L+\frac{v}{2\Omega}\right).
\end{equation}
In Fig.~\ref{y_pm} the first epicyclic maximum is at about 1500 pc. This is consistent with eq.~\ref{eq:yc+v} if we  assume that the average excess velocity is about 0.34 km/s.

\subsection{Milky-Way potential, circular orbit}\label{ssec:fgpco}
\begin{figure}
  \includegraphics[width=84mm]{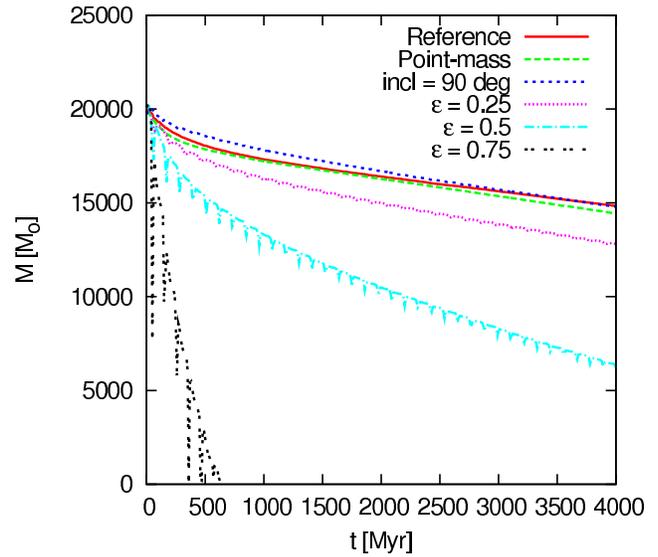}
  \caption{Mass evolution of the clusters at an initial galactocentric distance of 8.5 kpc.  Here $M$ is the mass inside radius $x_L$, determined using equation~\ref{eq:rtide}. The reference cluster is the most unperturbed and therefore survives for the longest time. The point-mass model compares well with the reference model since they have the same concentration and are both not subject to tidal variations. The cluster with the inclined orbit (Sec.~\ref{ssec:fgppo}) shows the slight influence of disk shocks, which are rather negligible at a galactocentric radius of 8.5 kpc for a cluster of the given concentration. On the contrary, eccentric orbits (Sec.~\ref{ssec:fgpeo}) significantly decrease a cluster's lifetime. Also visible from the figure is the fact that for all models mass-loss takes place rather smoothly and does not happen in steps. Moreover, the effect of the pericentre dips of the tidal radius on the bound mass can be seen, but it shows that most stars get recaptured by apocentre.}
  \label{M8500}
\end{figure}
\begin{figure}
\includegraphics[width=84mm]{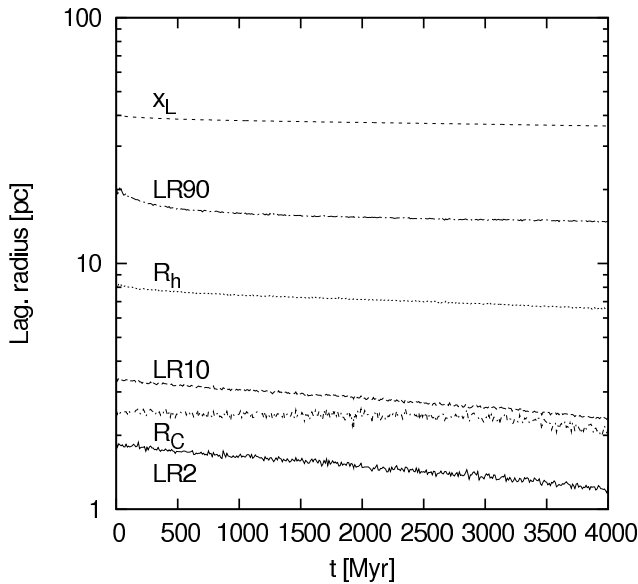}
  \caption{Evolution of the 2\%, 10\%, 50\% ($R_h$) and 90\% Lagrange radii of the reference cluster. Additionally, the tidal radius ($x_L$) and the core radius ($R_C$) are shown. The former shows the external/tidal evolution of the cluster which, in this case, is just affected by the mass-loss of the cluster itself, while the latter reflects the internal dynamical evolution of the cluster. By 4 Gyr the cluster is beginning to go into core collapse.}
  \label{LRref}
\end{figure}
\begin{figure}
\includegraphics[width=84mm]{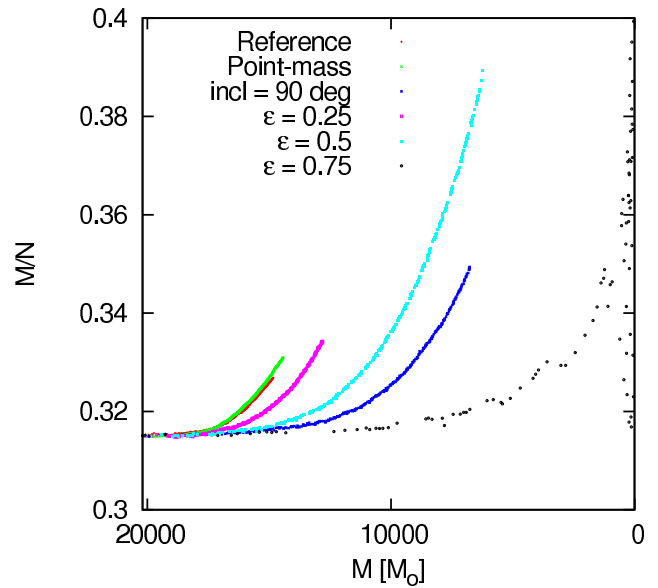}
  \caption{The average mass of stars, $M/N$, versus bound mass, $M$, for the clusters at 8.5 kpc. The preferential loss of low-mass stars is very pronounced in all clusters. Since preferential loss is a phenomenon of two-body relaxation, the most unperturbed clusters, i.e. the reference cluster and the cluster in the point-mass potential, show the earliest increase in $M/N$ with decreasing $M$. As all clusters, except the point-mass case, have the same initial relaxation time, the start of mass segregation gives a quantification of the importance of two-body relaxation compared to the tidally induced mass loss. Computations were halted after 4 Gyr so not all clusters dissolve completely.}
  \label{MN8500}
\end{figure}
\begin{figure}
\includegraphics[width=84mm]{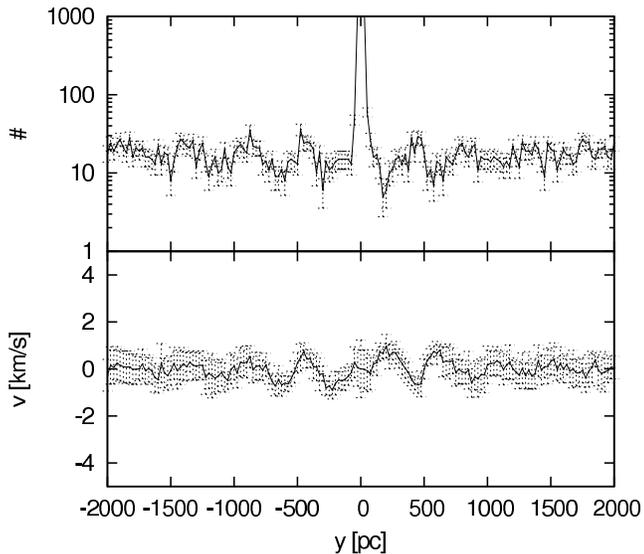}
  \caption{Number distribution of stars along the tidal tails in bins of 25 pc (upper panel) and corresponding mean velocities of the stars within the bins  (with respect to the orbital velocity of the cluster, eq.~\ref{eq:v}) for the reference cluster (discussed in Sec. \ref{ssec:fgpco}). The snapshot was taken when the cluster had a mass of about $16400 \msun$ at $t=2.0$ Gyr, which means a theoretical tidal radius of roughly 37 pc and a distance of the first-order overdensity of about 350 pc (equation \ref{eq:yc}). Error bars in the upper panel are Poisson errors and in the lower panel show the velocity dispersion within the corresponding bin (eq.~\ref{eq:sigma}).
  }
  \label{y_ref}
\end{figure}
\begin{figure*}
\includegraphics[width=168mm]{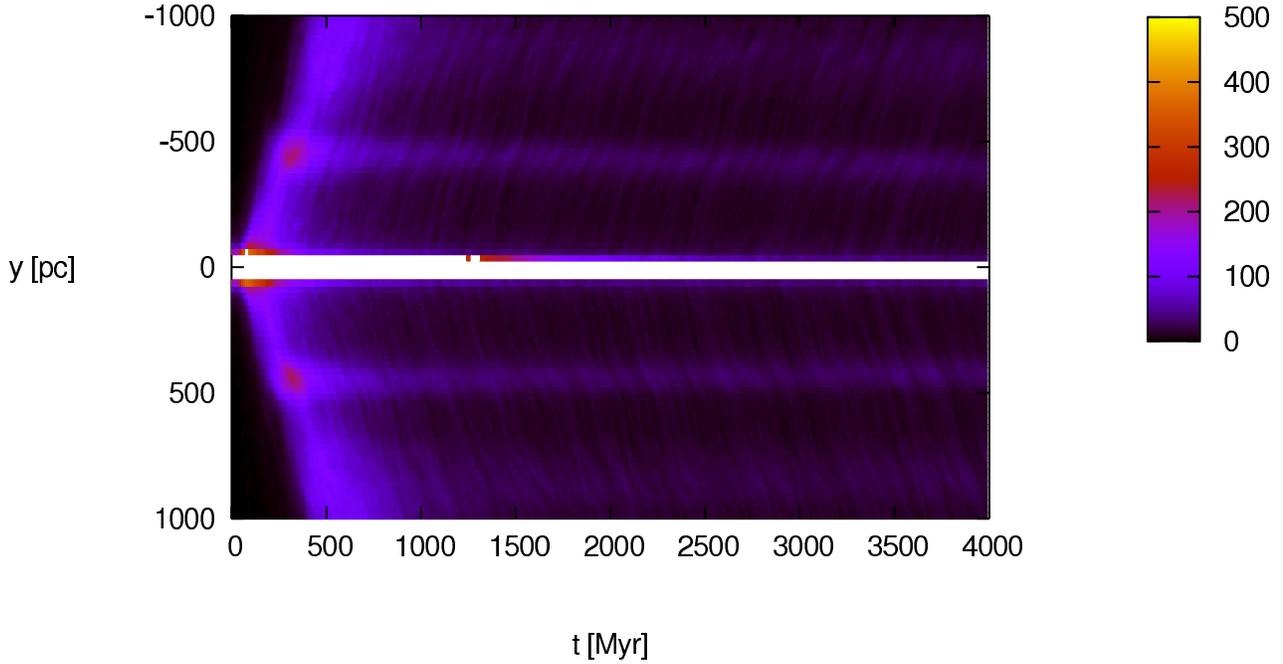}
  \caption{Time evolution of the number distribution of stars along the tidal tails in bins of 25 pc for the reference cluster (discussed in Sec. \ref{ssec:fgpco}). The first- and second-order overdensities can be seen at about $\pm$400 pc and $\pm$800 pc, respectively, decreasing slightly in distance as the cluster loses mass and the tidal radius decreases. The color bar on the right shows the color coding of the number of stars per bin.}
  \label{yc_ref}
\end{figure*}
The cluster described in this section will be the reference for all following clusters, since all of them are moving in the same galactic potential but on more complex orbits. Here we discuss some basic evolutionary parameters of this cluster (i.e. bound mass, Lagrange radii and mean stellar mass) and only point out significant differences for the other clusters in the following sections. The aim is to understand the evolution of the reference cluster in detail before we look at the tidal tails and the overdensities therein.

\subsubsection{Internal evolution}
Putting the cluster of Section \ref{ssec:pmgco} into a Milky-Way potential will not significantly change its evolution as long as it still moves on a circular orbit in the disk and with the same angular velocity. The only change is the steepness of the potential which, in the formalism described above, alters the epicyclic frequency, $\kappa^2$, and therefore changes the value of the tidal radius. Since the Milky-Way potential in the disk falls off less steeply, $\partial^2\Phi/\partial R^2$ is smaller and the tidal radius therefore larger.

This cluster is not perturbed by a time-dependent tidal field. Hence, as no external energy comes in, its lifetime is longest (Fig \ref{M8500}). During the first 500 Myr there is a small burst of primordial escapers. Thereafter, the mass-loss rate is about constant. After four Gyr the cluster still has about 75\% of its initial mass and is only just beginning to go into core collapse (Fig.~\ref{LRref}). The core radius, $R_C$, in this figure shows the internal dynamical evolution of the reference cluster. It is calculated using the algorithm (cf. \citet{Casertano85})
\begin{equation}
 R_C = \sqrt{\frac{\sum_i^N r_i^2\rho_i^2}{\sum_i^N\rho_i^2}},
\end{equation}
where $r_i$ is the distance of the $i$th star from the density centre and $\rho_i$  is the mass density around this star, computed using the distance to its fifth closest neighbour and the mass within this radius.

A good indicator for the formation of epicyclic overdensities is ongoing two-body relaxation in the cluster, as escape due to two-body relaxation ensures evaporative escape conditions (Sec.2.2). In Fig.~\ref{MN8500} the mean mass of bound stars versus bound mass is shown. Since a cluster is expected to lose preferentially low-mass stars, as those are most probably accelerated to velocities above the escape velocity, and this effect is due to energy equipartition (via two-body relaxation), Fig.~\ref{MN8500} gives a measure of the effectiveness of two-body relaxation for the escape of stars. The preferential loss is clearly visible in all clusters. We can see that the reference cluster, as it is the most unperturbed one, shows the earliest increase in mean mass with decreasing bound mass.

\subsubsection{Evolution of the tidal tails}
As we have seen that two-body relaxation plays a major role in the dissolution process, and escape due to two-body relaxation leads to epicyclic overdensities, the reference cluster also shows the expected density enhancements at the predicted positions (Fig.~\ref{y_ref}). As mentioned in Sec.~\ref{sec:rt}, we can use $\kappa = 1.4 \Omega$ for this orbit in the Milky-Way potential, which gives
\begin{eqnarray}\label{eq:y_cref}
 y_C &\simeq& \pm 3\pi x_L,\\
x_L^3 &\simeq& \frac{GM}{2\Omega^2}.
\end{eqnarray}
Hence, the epicycles of the escaping stars are about four times shorter than in the case of a point-mass galaxy while the tidal radius is almost 10\% larger. Still, the epicyclic approximation is correct, as can be seen in Fig.~\ref{y_ref}, where the binned stars along the tidal tails are shown at 2.0 Gyr when the cluster has a bound mass of about $16400 \msun$. The corresponding tidal radius and the predicted position of the first overdensity, $y_C$, are given in the figure caption.  As in Fig.~\ref{y_pm} the first-order epicyclic maximum is at a somewhat larger distance than the predicted $y_C$. The difference is about 50-100 pc which, using eq.~\ref{eq:yc+v}, corresponds to an average excess velocity of about 0.27-0.55 km/s.

The lower panel of Fig.~\ref{y_ref} also shows the predicted velocity variations. Moreover, the average orbital velocity with respect to the cluster is again in good agreement with the predictions of eq.~\ref{eq:vccorr}, which yields an average orbital velocity of about $\overline{v} = 0$ kms$^{-1}$ for the given parameters. This average value is intuitively expected, since the rotation curve of the assumed galactic potential is almost flat at this galactocentric radius. The role of the flat rotation curve can be nicely seen by comparing Fig.~\ref{y_ref} with Fig.~\ref{y_pm} where larger velocity differences between the leading and trailing tidal tails are seen.

When comparing Fig.~\ref{y_ref} with Fig.~\ref{y_pm}, which show similar clusters at the same dynamical state, i.e. at a bound mass of about $16000 \msun$, but in a different tidal field, it should be noticed that, besides the fact that $y_C$ is approximately four times larger in Fig.~\ref{y_pm}, the overdensities are also far more extended in Fig.~\ref{y_pm}, also by a factor of about four. Furthermore, the peak density outside the cluster is larger by a factor of two in the case of the more realistic reference cluster. This is due to the mean drift velocity of the stars along the tidal tails (eq.~\ref{eq:vc}). Comparing $v_C$ for the two clusters we see that, while the escaping stars of the cluster in the point-mass potential have a mean drift velocity of about 5 kms$^{-1}$, the tail stars of the reference cluster move with only about 2 kms$^{-1}$ along the tail, i.e. the tail is denser on average because the mass-loss rate of both clusters is approximately the same (see Fig.~\ref{M8500}). This implies that the tails of the reference cluster, and especially its overdensities, would be much easier to observe.

Fig.~\ref{yc_ref} shows a time series of the number distribution along the tidal tails. In this figure the cluster is omitted to allow a larger dynamical range in the representation of the number density of the tails. The first- and second-order overdensities can be seen with distances decreasing slightly with time as the mass, and hence the tidal radius, of the system monotonically decreases. There is also an initial transient overdensity, which propagates rapidly away from the cluster at small times. This is due the initial presence of ``primordial escapers'' (Sec.3.1), and is an artefact of the initial conditions.  But it is an interesting feature, as it illustrates how very differently the overdensities evolve if they are due to a single pulse of mass loss.

\subsection{Milky-Way potential, eccentric orbits}\label{ssec:fgpeo}
\begin{figure*}
\includegraphics[width=168mm]{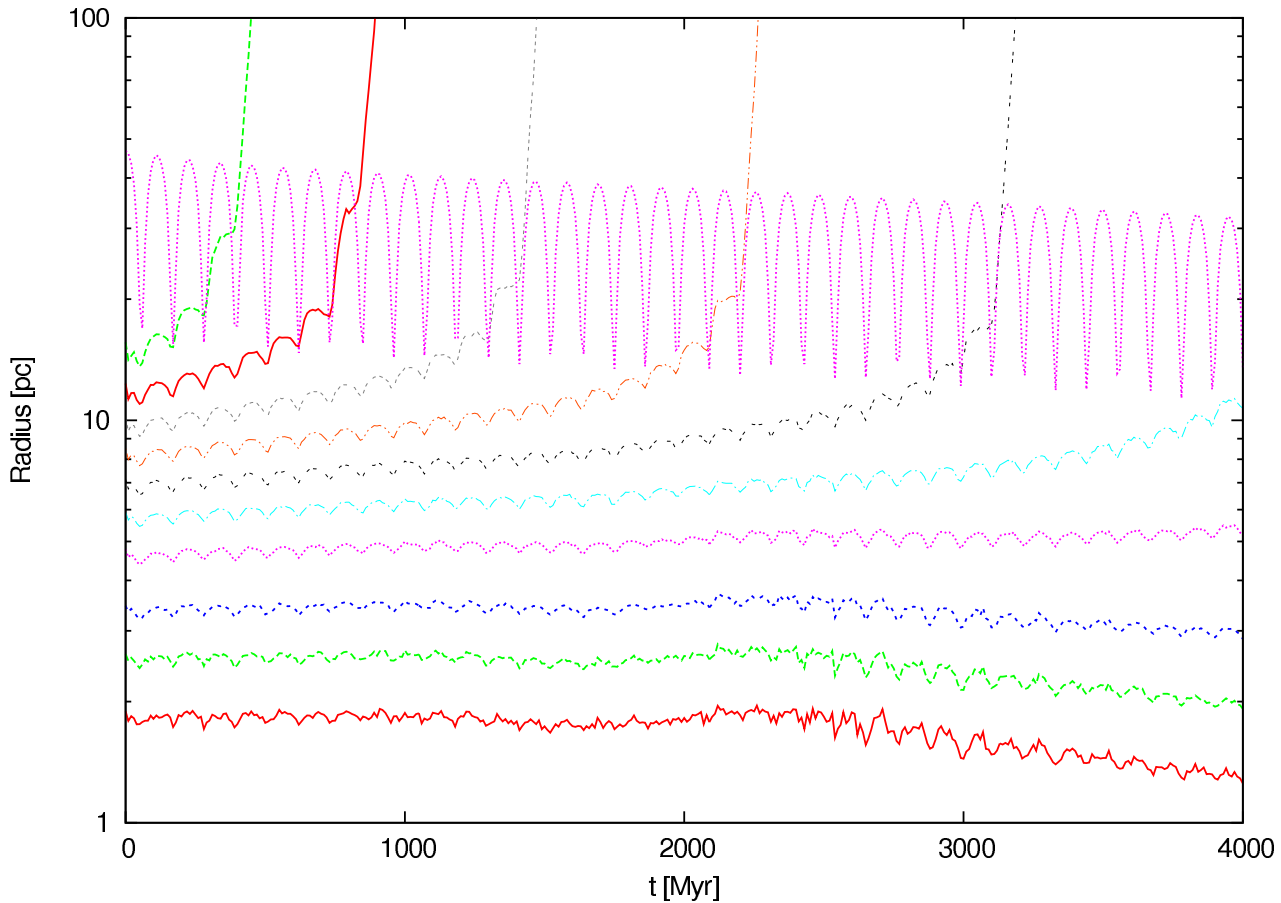}
  \caption{Time evolution of the mass shells (Lagrange radii) of the cluster with an eccentricity of 0.5 and an apocentre distance of 8.5 kpc. The lines show the radii containing 2, 5, 10, 20, 30, 40, 50, 60, 70, 80 and 90 percent of the initial cluster mass (about 20000 $\msun$). The periodically changing line, which is uppermost at $t = 0$, is the tidal radius. The periodic expansion and compression of the whole cluster through the pericentre passages can be observed in all shells. Only the innermost shells show a progressive contraction while the other shells show an accelerating expansion. If a shell gets unbound during a pericentre passage, the shell will quickly expand and move away from the cluster. If the shell can get recaptured by the tidal radius, the expansion will be slowed down or even reversed.}
  \label{1224KR}
\end{figure*}
\begin{table}
\begin{minipage}{84mm}
\centering
 \caption{Parameters for the formalism described in Sec.~\ref{ssec:yc of t} for the models at 8.5 kpc distance. The first two columns are taken from Table~\ref{table1}, whereas $\overline{R}$ gives the mean galactocentric radius, $\overline{V}$ is the mean orbital velocity of the cluster and $\Delta t$ is the estimate of the time difference between the cluster and the first-order overdensity reaching the same phase of the orbit on their orbit about the galaxy as derived in Section~\ref{ssec:yc of t}.
}
\label{table2}
\begin{tabular}{ccccc}
\hline
   $incl$ [deg]& $\epsilon$ & $\overline{R}$ [kpc]& $\overline{V}$ [kms$^{-1}$]& $\Delta t$ [Myr]\\
\hline
0 & 0.00 & 8.50 & 220 & 1.72\\
0 & 0.25 & 6.99 & 214 & 1.58\\
0 & 0.50 & 6.22 & 202 & 1.61\\
0 & 0.75 & 5.94 & 188 & 1.76\\
90& 0.04 & 8.93 & 204 & 2.01
\end{tabular}
\end{minipage}
\end{table}
\begin{figure}
\includegraphics[width=84mm]{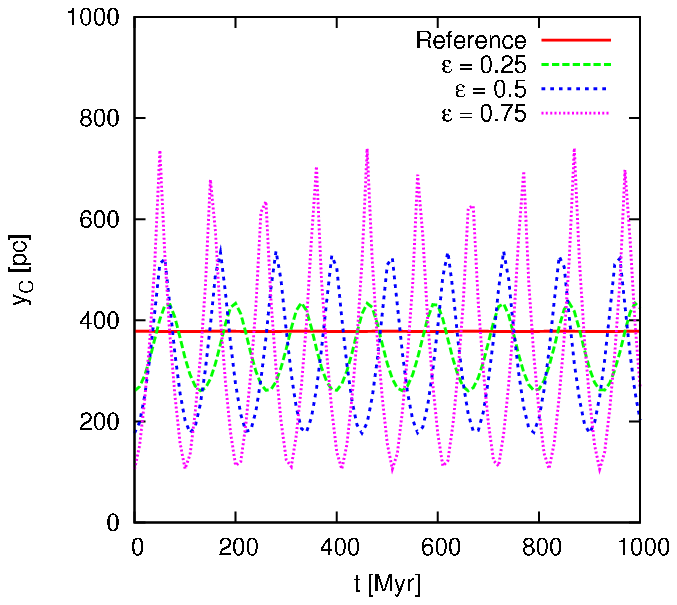}
  \caption{Distance of the first-order epicyclic overdensity, $y_C$, estimated with the formalism described in Sec.~\ref{ssec:yc of t} for the eccentric orbits at 8.5 kpc. The corresponding values of $\overline{R}$,  $\overline{V}$ and $\Delta t$ can be found in Table~\ref{table2}. The values for the velocity $V$ and the acceleration $A$ on the cluster were extracted from our computations in time steps of 10 Myr. For comparison the value of the reference cluster is also shown. On eccentric orbits the distance $y_C$ starts oscillating about a mean value, where the amplitude of the oscillations increases with increasing eccentricity. The mass of the cluster in this plot is fixed to the initial mass of 20000 $\msun$. If mass loss was taken into account the mean value as well as the amplitude of the oscillations would decrease with time. The predictions made within this figure can be compared with the $N$-body results in Fig.~\ref{yc_e025} and Fig.~\ref{yc_e05}. (We can also see from this figure that, unlike in a Keplerian potential, the period of the orbits decreases with increasing eccentricity. This further increases the effect of higher eccentricities on the internal cluster evolution.)}
  \label{yct_ecc}
\end{figure}
\begin{figure*}
\includegraphics[width=168mm]{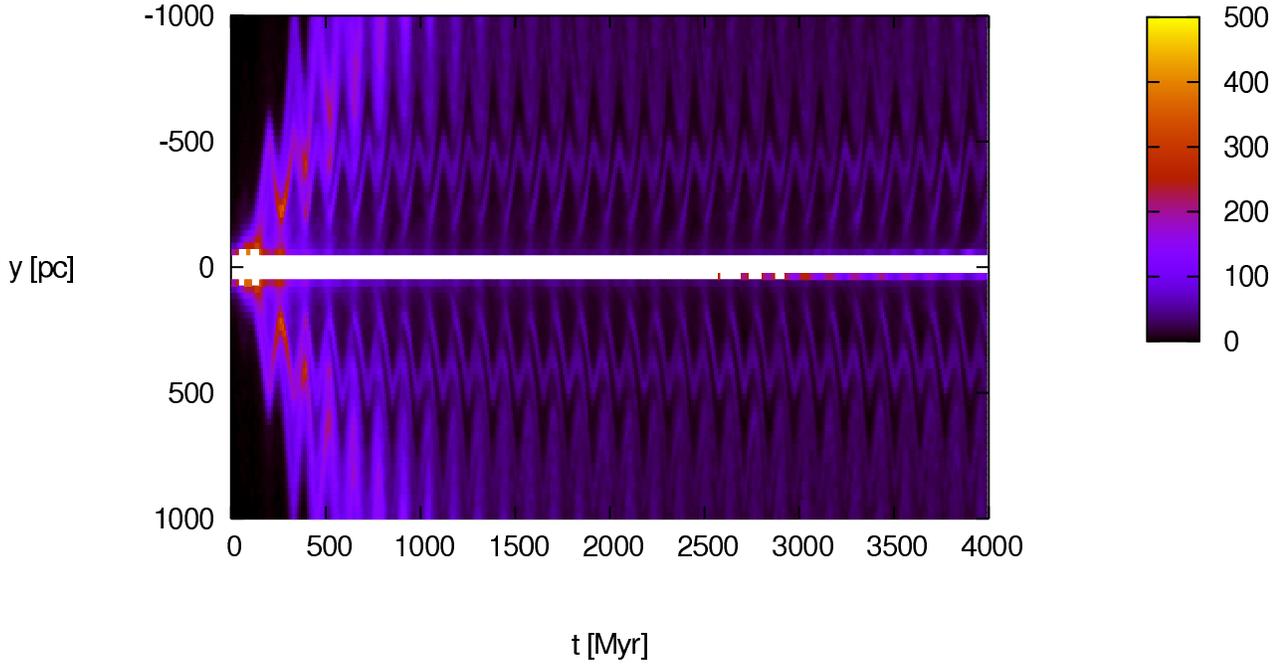}
8  \caption{As Fig.~\ref{yc_ref}: time evolution of the number distribution of stars along the tidal tails in bins of 25 pc for the cluster with apogalactic distance of 8.5 kpc and orbital eccentricity of 0.25 (discussed in Sec. \ref{ssec:fgpco}). The first- and second-order overdensities can be seen, periodically changing in distance but in mean following the trend of the reference cluster (cf. Fig.~\ref{yc_ref}). The computations agree with our predictions made in Sec.~\ref{ssec:yc of t} fairly well, as can be seen when compared with Fig.~\ref{yct_ecc}.}
  \label{yc_e025}
\end{figure*}
\begin{figure*}
\includegraphics[width=168mm]{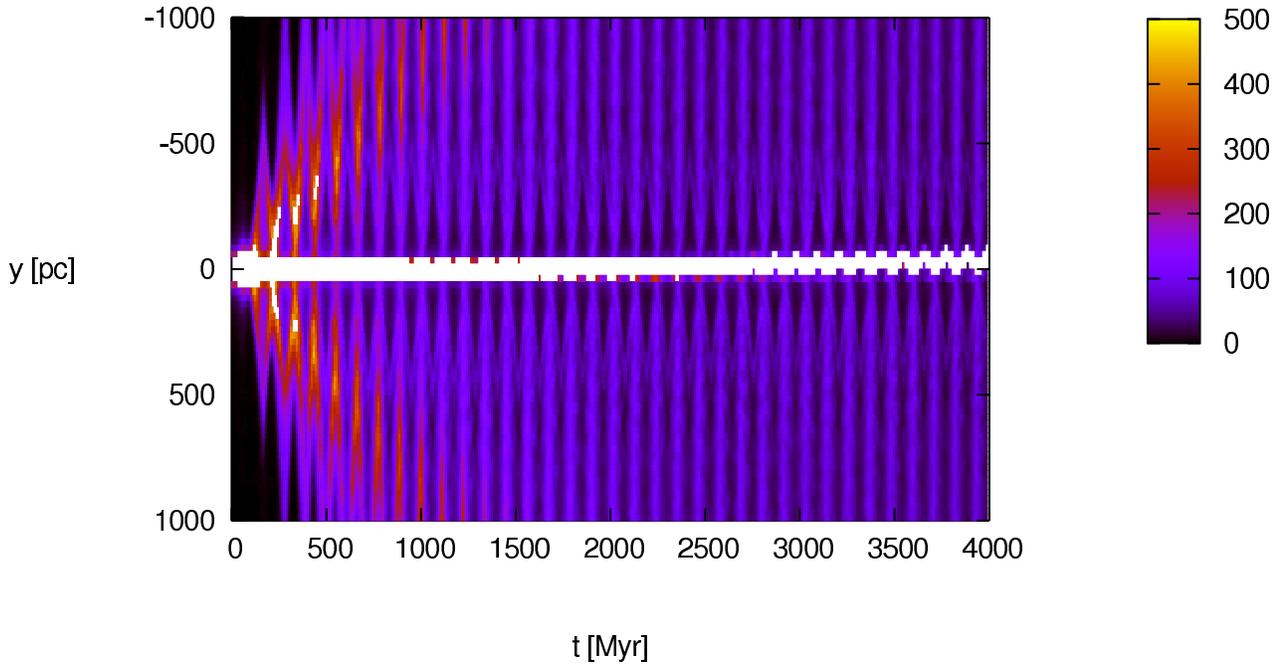}
  \caption{As Fig.~\ref{yc_ref}: time evolution of the number distribution of stars along the tidal tails in bins of 25 pc for the cluster with apogalactic distance of 8.5 kpc and orbital eccentricity of 0.5 (discussed in Sec. \ref{ssec:fgpco}). The first-order overdensities can be seen especially in perigalacticon as a faint blue structure, following our predictions made in Sec.~\ref{ssec:yc of t} fairly well (cf. Fig.~\ref{yct_ecc}). The most prominent overdensity at early times, however, is due to primordial escapers. It can be followed as a red structure moving slowly away from the cluster.}
  \label{yc_e05}
\end{figure*}
\begin{figure*}
   \includegraphics[width=164mm]{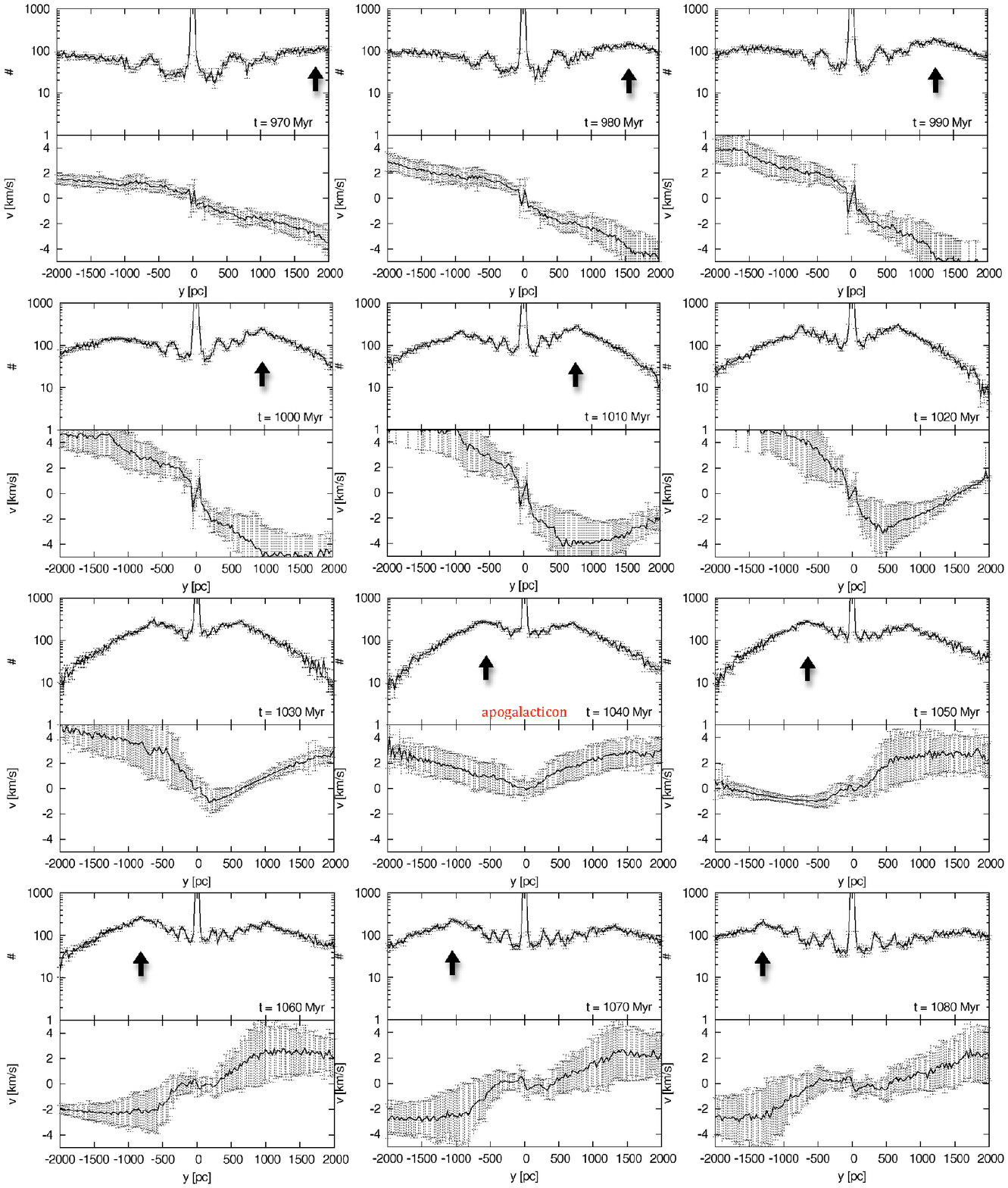}
  \caption{Number distribution of stars along the tidal tails in bins of 25 pc (upper panels) and corresponding mean velocities of the stars within the bins (with respect to the orbital velocity of the cluster, eq.~\ref{eq:v}) for the cluster with an eccentricity of 0.5 and an apogalactic distance of 12.75 kpc (discussed in Sec. \ref{ssec:fgprgal}). The twelve consecutive snapshots were taken when the cluster had a mass of about $14000 \msun$ starting at $t = 970$ Myr in time steps of 10 Myr. A full revolution takes about 170 Myr. So, the first panel is 20 Myr after the cluster has been at perigalacticon where the tails where maximally stretched, in the middle panel of the third row ($t=1040$ Myr) the cluster is in apogalacticon and maximally compressed. In the right panel of the forth row the cluster is 40 Myr away from perigalacticon. In this time series the difference in average number of stars per bin between peri- and apogalacticon is observable which exceeds a factor of three (compare, for example, the left panels in the first and third row in the vicinity of the cluster). Furthermore, the asymmetric compression can be seen, wandering from positive to negative y-values (indicated by the arrow), as an asymmetric density distribution when the leading and the trailing tail are carefully compared. The part of the tails which is compressed most shows a number density which is about twice as high as the same part in the opposite tail. The asymmetric acceleration and deceleration of the tail stars can also be seen in the velocity distributions, where the velocity dispersion is large in parts of the tails which are broad and small in parts which are thin.}
  \label{y_e075}
\end{figure*}

We study the influence of pericentre passages on the formation of tidal tails by using eccentric orbits, but remain within the disk in order to avoid disk shocks. The eccentricities, $\epsilon$, of the chosen orbits are 0.25, 0.5 and 0.75, where
\begin{equation}\label{eq:ecc}
\epsilon = \frac{R_{apo}-R_{peri}}{R_{apo}+R_{peri}}.
\end{equation}
The clusters start at a radius of 8.5 kpc and have an initial velocity which is reduced by a corresponding amount compared to the circular velocity, and hence 8.5 kpc is their apocentre distance. The corresponding values of $R_{peri}$ can be found in Table \ref{table1}.

\subsubsection{Internal evolution}
The clusters computed in this way dissolve faster than the reference cluster (Fig.~\ref{M8500}), which is expected, because the more eccentric the orbit the stronger is the average tidal field \citep{Baumgardt03}. Furthermore, for an eccentricity of 0.25 there are already small periodic wobbles, which are  visible in Fig.~\ref{M8500} and which grow with increasing $\epsilon$. These are due to the periodically changing value of the tidal radius  $x_L$, which  is smallest at pericentre,  and this sets loose a fraction of stars, of which  most are recaptured by apocentre. As discussed in Section \ref{ssec:delay}, due to this recapture, stars are not lost in a pulse during a pericentre passage: the mass curve rather follows a smooth line (if the pericentre dips are neglected), and not a series of steps.

This behaviour, in which we follow the mass within the tidal radius calculated using eq.~\ref{eq:rtide}, is very different from the behaviour found by \citet{Dehnen04} (their fig.~4), who included only stars lying within the initial tidal radius  for the computation of internal quantities of the cluster, such as the total mass. Their approach suggests that stars are lost in pulses during  pericentre passages or disk shocks, which is in strong contrast with our finding. Differences in the initial conditions, orbital parameters, galactic model and even computational method all have a role in this comparison. We note, however, that recent results (K\"upper, Kroupa, Baumgardt \& Heggie, in preparation) show that for computations of internal quantities it is more appropriate to use the apogalactic tidal radius or the mean tidal radius mentioned in Sec.~\ref{ssec:yc of t}.

While the strength of these dips in the mass curve do not reflect permanent escape, they do indicate what fraction of stars is seriously affected by a pericentre passage or a disk shock, since it shows how many stars are at large radii and get temporarily unbound.  (Those stars which are most affected by a pericentre passage or a disk shock are those at large radii (cf. equations \ref{eq:dE} and \ref{eq:dE2})). But, as we can see in these dips, not all stars that have been once outside the tidal radius are lost from the cluster. As mentioned in Sec.~\ref{ssec:delay}, the loss of these outlying stars cannot happen instantaneously, essentially because their orbital time about the cluster is on the order of the orbital time of the cluster about the galactic centre.

The finite breadth of the dips (e.g. for $\varepsilon = 0.25$) also tells us that not all eccentricities lead to tidal shocks in a classical sense. Therefore, we have tried to avoid using the term \textit{shock} in connection with eccentric orbits, and speak of the more general \textit{pericentre passages}, which implies a broader range of time scales.

The  behaviour we are describing can be observed in Fig.~\ref{1224KR}, where the time evolution of mass shells of the cluster with an eccentricity of 0.5 and an apocentre distance of 8.5 kpc is shown. The figure demonstrates the periodic expansion and compression of the whole cluster, but the amplitude is much smaller than the change in tidal radius. Superimposed on this externally induced oscillation is the internal dynamical evolution of the cluster. While the innermost shells are contracting due to mass segregation, the remaining shells are expanding.

If a shell gets unbound during a pericentre passage it will expand thereafter, where the amount of expansion depends on the time the shell spends outside the tidal radius. If this time is short (e.g. the curve for 80\% in Fig.~\ref{1224KR} at about 750~Myr, the expansion will be slow enough that the shell can be recaptured by the growing tidal radius when the cluster moves from peri- to apocentre. A recapture will decelerate the expansion or even reverse it. When no recapture is possible, the shell will quickly move away from the cluster vicinity within a fraction of the cluster orbital period. Here again it can be seen that escape happens throughout the whole orbit and is not heavily concentrated at perigalacticon.

But how does the eccentricity of the orbit affect two-body relaxation in the cluster and what is the effect on the escape conditions? Fig.~\ref{MN8500} shows that preferential loss of low-mass stars is increasingly suppressed compared to the reference cluster with increasing eccentricity, since the fraction of lost stars with arbitrary masses increases. Nevertheless, all clusters show an increasing mean stellar mass, such that two-body relaxation is evidently at work. Provided that the escape conditions for a majority of stars are still evaporative, and the eccentricity of the orbits does not induce too large a large scatter in escape conditions, we may still expect that epicyclic overdensities will arise.  We now examine the tidal tails to check this expectation.

\subsubsection{Evolution of the tidal tails}
The tidal tails of clusters on eccentric orbits behave differently from those of clusters on circular orbits in several ways. First of all, the coordinate system in which the cluster is at rest is no longer uniformly accelerated, and therefore the epicycle approximation cannot be applied without modification. For this reason, the solutions of the equations of motion of KMH (therein eq. 4-6) do not hold any more. Hence, there is no simple relation between the tidal radius and the distance of the density maxima unless we make further assumptions (Sec.~\ref{ssec:yc of t}). Nevertheless, the equation for the tidal radius (eq. \ref{eq:rtide}) is still valid since the orbits lie in the plane of symmetry of the potential and therefore the angular momentum is conserved. But the value of $\partial^2\Phi/\partial R^2$ now changes with time as the galactocentric distance oscillates, and the angular velocity, of course, also varies with time. This gives a variation of the tidal radius (between peri- and apocentre)  of order 90\% of the apogalactic value for $\epsilon = 0.75$  for the clusters with an apogalactic distance of 8.5 kpc.

\subsubsection*{Epicyclic overdensities}
As we  expected from theoretical considerations above, we also find epicyclic overdensities in the tidal tails of clusters on eccentric orbits.  Indeed, all clusters in the range of eccentricities we studied show these overdensities. The epicyclic origin of the  overdensities we have found can be seen by looking at time series of the number distribution along the tidal tails, as the distance $y_C$ should be related to the tidal radius.

Since the tidal radius is large at apocentre and small at pericentre, the density maxima are expected at a larger distance at apogalacticon and vice versa. But the opposite is the case, which is due to the non-linearly accelerated rest frame: from  apo- to perigalacticon the cluster and its tails get accelerated and stretched, while from peri- to apocentre the tails get compressed as the the whole system is decelerated (cf. \citealt{Piatek95}). Hence, at pericentre the cluster is fastest and the distance between the maxima is largest, whereas the maxima and the whole tails get most compressed at apogalacticon - which is also the explanation why the tidal tails of Palomar~5 are so clearly visible since it is near the apocentre of its orbit about the Galaxy \citep{Odenkirchen03}.

If we apply the assumptions of Sec.~\ref{ssec:yc of t} we can predict the positions of the overdensities even for the eccentric orbits. Therefore we average the galactocentric radius and the orbital velocity of the cluster for each model in Table~\ref{table2} and estimate a time difference, $\Delta t$, by which the cluster and the first-order overdensity should be separated in reaching the same phase of the orbit. The variations of $y_C$ for the different orbital types at a fixed mass of 20000~$\msun$ are shown in Fig.~\ref{yct_ecc} for several revolutions.

Obviously, the distance $y_C$ now strongly depends on the position of the cluster on its orbit; from apo- to pericentre it varies increasingly with increasing eccentricity. This periodic change of $y_C$ is also observable in Fig.~\ref{yc_e025} where the change of the number distribution along the tidal tails with time is shown for the cluster at an apogalactic distance of 8.5 kpc and an eccentricity of 0.25. While the reference cluster (Fig.~\ref{yc_ref}) shows the predicted monotonic decrease of $y_C$, the clusters on eccentric orbits have an orbital variation imprinted on this monotonic decrease. Our estimate from Sec.~\ref{ssec:yc of t} shows good agreement with the simulations; both vary by a factor of 1.7 about a mean value which is lower than the corresponding value of the reference cluster. The same holds for the cluster with an orbital eccentricity of 0.5, shown in Fig.~\ref{yc_e05}. The epicyclic overdensities in this figure are quite faint compared to the initial mass loss, which is due to primordial escapers. Nevertheless, the location of the first-order epicyclic overdensity is in good agreement with  the predictions made using the formalism of Sec.~\ref{ssec:yc of t}.\footnote[2]{For the cluster with an orbital eccentricity of 0.75 the time resolution of our output data is too poor compared with the orbital time. Moreover, the average density within the tails differs so much between apo- and perigalacticon such that it is hard to convincingly show the epicyclic overdensities in this kind of representation. But this is just a problem of representation.}, supporting the assumptions made for this purpose.

While  $y_C$ in Fig.~\ref{yct_ecc} varies for an eccentricity of 0.25 by a factor of about 1.5 between maximal and minimal distance from the cluster centre, and for an eccentricity of 0.5 by a factor of 2.7, the amplitude of variation for $\epsilon = 0.75$ is even a factor of 7.5. Moreover, for the latter eccentricity the compression at apogalacticon is so strong that the overdensities are literally pushed back into the cluster such that they overlap with each other and the cluster and nearly disappear in our number distribution plots.

\subsubsection*{Asymmetry overdensities}
The non-uniform acceleration along the orbit also leads to an asymmetry in extended tidal tails, since in the case when the cluster is exactly at apocentre, for example, the leading tail is already accelerating towards the galactic centre while the trailing tail is still slowing down. The difference in acceleration results in an asymmetry in compression between the leading and the trailing tail, and thus in an asymmetric stellar density within the two tails. The asymmetry increases with a larger eccentricity and is also visible in the velocity distribution along the tidal tails.

Nevertheless, the epicyclic overdensities stay visible throughout the whole orbital period. We demonstrate this for the cluster with an orbital eccentricity of 0.5 but at an apogalactic distance of 12.75 kpc. We do this because we generated outputs of \textsc{NBODY4} every 10 Myr throughout our computations, and for clusters at smaller apogalactic distances the time resolution is too poor for a convincing demonstration. For this cluster with $\epsilon = 0.5$ the orbital time is about 170 Myr, we therefore show 12 consecutive snapshots starting at 970 Myr in steps of 10 Myr in Fig.~\ref{y_e075}. At that time it has a mass of about $14000\msun$ and is first close to pericentre, then at 1040 Myr in apocentre and in the last panel, at 1080 Myr, on its way to pericentre again.

Not only is the compression significant, increasing the stellar density within the tails by, at least, a factor of three at apogalacticon compared to perigalacticon, but also the asymmetry is clearly visible and can be followed throughout the snapshots wandering from right to left (indicated by a black arrow) and enhancing the number density by a factor of about two compared to the same part in the opposite tail. This asymmetry may be interpreted as another overdensity type, which may be very helpful for determining the orbit of an observed cluster like Pal~5.

The most important point about the studied pericentre passages is that all observed overdensities are due to epicyclic motion of escaping stars or an asymmetric acceleration of the tidal tails. There is no detectable overdensity being obviously the result of a pericentre passage (if we neglect the loss of the primordial escapers during the first revolution). Hence, even for the most eccentric orbits studied here, the conditions of slow escape are well fulfilled for a majority of escapers such that epicyclic overdensities can be observed. This further supports the assumptions made in Sec.~\ref{ssec:yc of t} which will be further investigated in K\"upper, Kroupa, Baumgardt \& Heggie (in preparation).

\subsection{Milky Way potential, inclined orbit}\label{ssec:fgppo}
\begin{figure*}
\includegraphics[width=168mm]{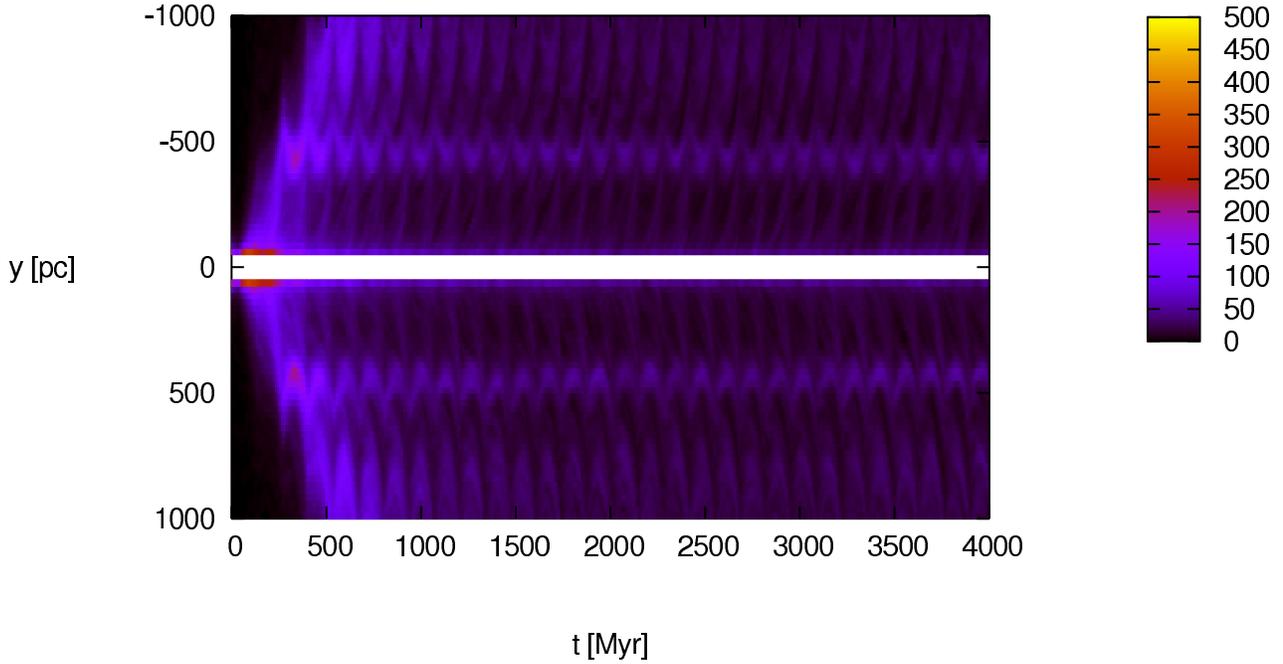}
  \caption{As Fig.~\ref{yc_ref}: time evolution of the number distribution of stars along the tidal tails in bins of 25 pc for the cluster with galactocentric distance of 8.5 kpc on an orbit with 90 deg inclination to the galactic disk (discussed in Sec. \ref{ssec:fgppo}). The first- and second-order overdensities can be seen, periodically changing in distance, as the orbit is slightly eccentric. But in mean the evolution is following the trend of the reference cluster (cf. Fig.~\ref{yc_ref}). Furthermore, there is no indication that a disk shock creates an overdensity by a temporary increase in the escape rate.}
  \label{yc_po}
\end{figure*}
\begin{figure}
\includegraphics[width=84mm]{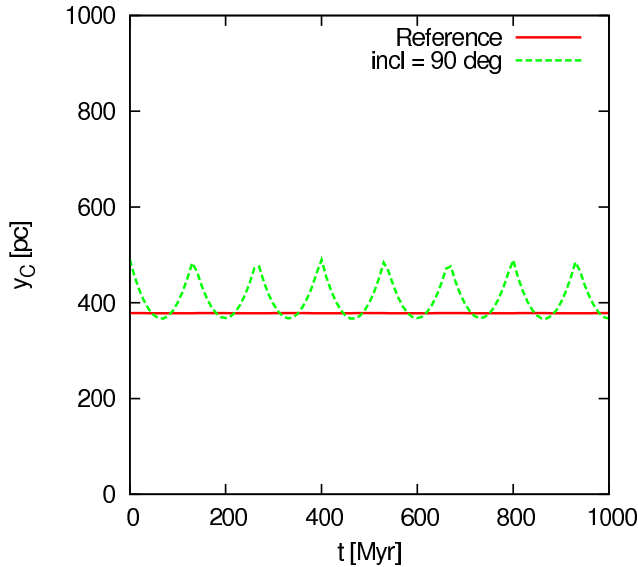}
  \caption{As Fig.~\ref{yct_ecc}: distance of the first-order epicyclic overdensity, $y_C$, estimated with the formalism described in Sec.~\ref{ssec:yc of t} for the inclined orbit at 8.5 kpc. The corresponding values of $\overline{R}$,  $\overline{V}$ and $\Delta t$ can be found in Table~\ref{table2}.  The values for the velocity $V$ and the acceleration $A$ on the cluster were extracted from our computations in time steps of 10 Myr. For comparison the value of the reference cluster is also shown. Like on eccentric orbits the distance $y_C$ starts oscillating about a mean value as a result of the acceleration and deceleration due to the galactic disk.  The mass of the cluster in this plot is fixed to the initial mass of 20000 $\msun$. If mass loss was taken into account the mean value as well as the amplitude of the oscillations would decrease with time. The predictions made within this figure can be compared with the $N$-body results in Fig.~\ref{yc_po}.}
  \label{yct_disk}
\end{figure}
We study disk shocks using a cluster on a circular orbit with 90 degree inclination to the disk. The tidal field which this cluster experiences is essentially the same as the one of the reference cluster, except for two disk shocks per revolution which put additional energy into the cluster according to equations \ref{eq:dE} \& \ref{eq:dE2}.

In Fig.~\ref{M8500} we can see that the effect of disk shocks at 8.5~kpc on the mass evolution of a cluster of the given concentration is not significant, in agreement with results from \citet{Vesperini97} and \citet{Gnedin99}.

Similar as in the reference cluster, the factor $\partial^2\Phi/\partial R^2$ (which is important for the calculation of the tidal radius but has nothing to do with the shock itself as it only reflects the potential change in the radial direction) is as good as constant within one orbit and just the angular velocity changes slightly when the cluster is accelerated or decelerated by the disk. This results in the orbit being slightly non-circular due to the non-spherical potential. The effective eccentricity is only about 0.04, though (see Table~\ref{table1}). This small non-zero eccentricity has a negligible effect on the evolution of the cluster.

The internal evolution of the cluster is also quite similar to the reference cluster (Fig.~\ref{MN8500}). Hence, epicyclic overdensities are observable in the number distribution of stars along the tidal tails, as can be seen in Fig.~\ref{yc_po} where the evolution of the tidal tails with time is shown. As the disk periodically accelerates and decelerates the cluster quite similarly to an eccentric orbit, the overdensities and the overall shape of the tails also get slightly compressed and stretched.

With the formalism which we introduced in Sec.~\ref{ssec:yc of t} and the values for the mean galactocentric radius and the mean orbital velocity we can predict the behaviour of $y_C$ for this model. The result is shown in Fig.~\ref{yct_disk} and is in satisfactory agreement with the $N$-body results as shown in Fig.~\ref{yc_po}.

Important about this model is that no overdensity due to a disk shock can be observed within the tails, i.e. there is no group of escapers moving along the tidal tail. All overdensities are due to epicyclic motion of escaping stars and mass loss due to a shock does not happen instantaneously with the shock, but happens delayed along the whole orbit as described in Section~\ref{ssec:delay}. Furthermore, disk shocks do not significantly increase the scatter in escape conditions for the epicyclic overdensities to vanish.

\subsection{Orbits at 4.25, 12.75 and 17 kpc}\label{ssec:fgprgal}
\begin{figure*}
\includegraphics[width=56mm]{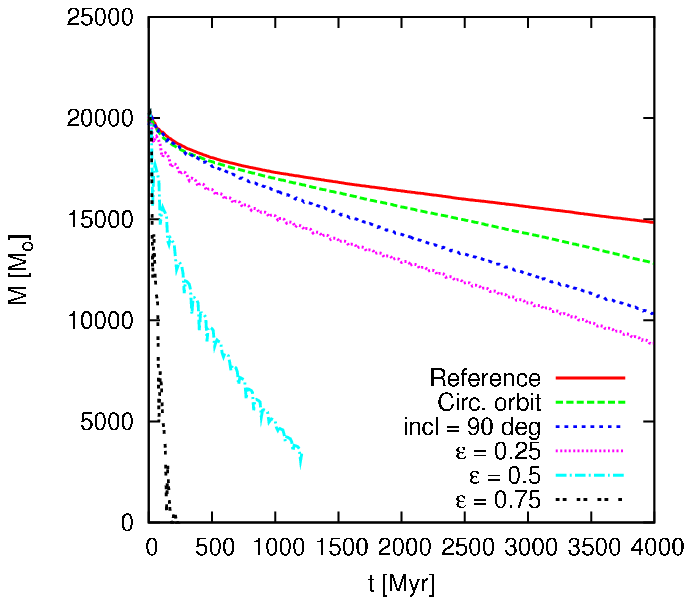}
\includegraphics[width=56mm]{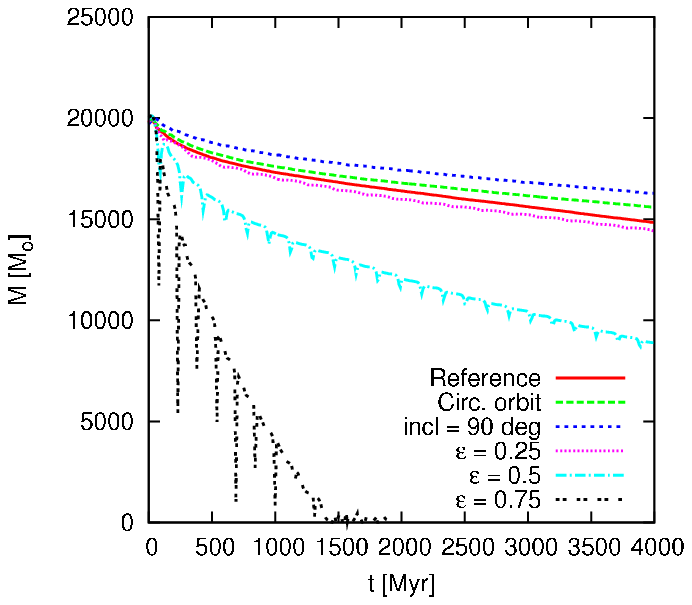}
\includegraphics[width=56mm]{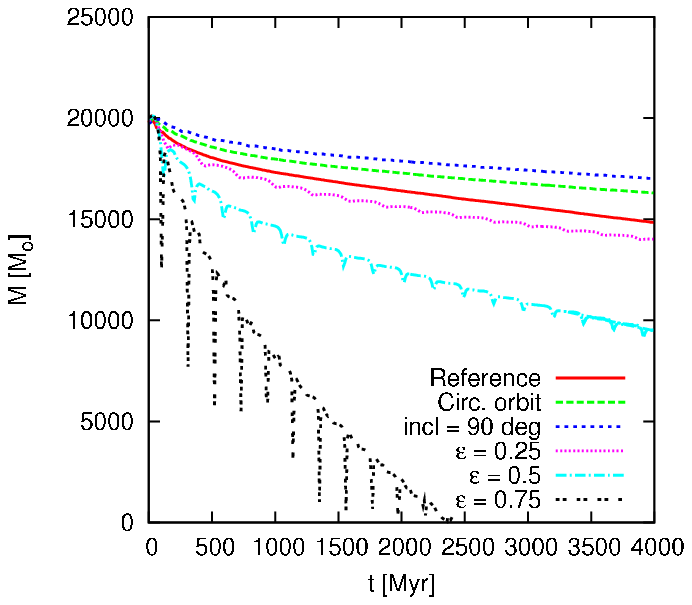}
  \caption{Mass evolution of the clusters at an initial galactocentric distance of 4.25 kpc (left), 12.75 kpc (middle) and 17 kpc (right). For comparison the reference cluster is also shown. The increase of importance of tidal influences with decreasing galactocentric radius is obvious, where disk shocks gain more influence compared to pericentre passages.}
  \label{Mrgal}
\end{figure*}
\begin{figure*}
\includegraphics[width=56mm]{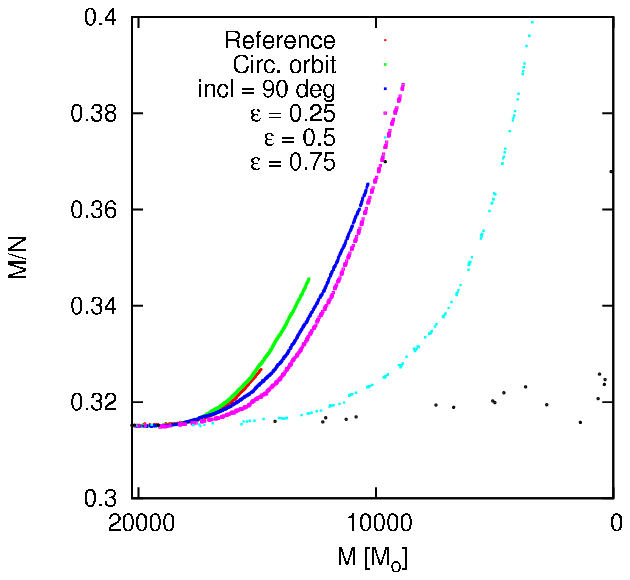}
\includegraphics[width=56mm]{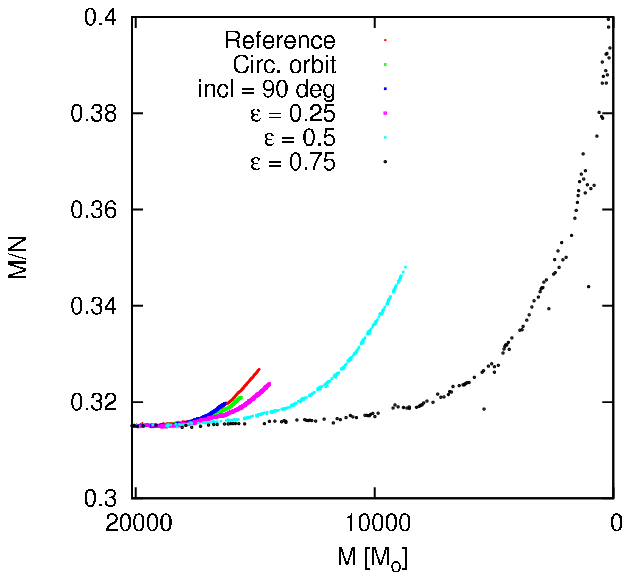}
\includegraphics[width=56mm]{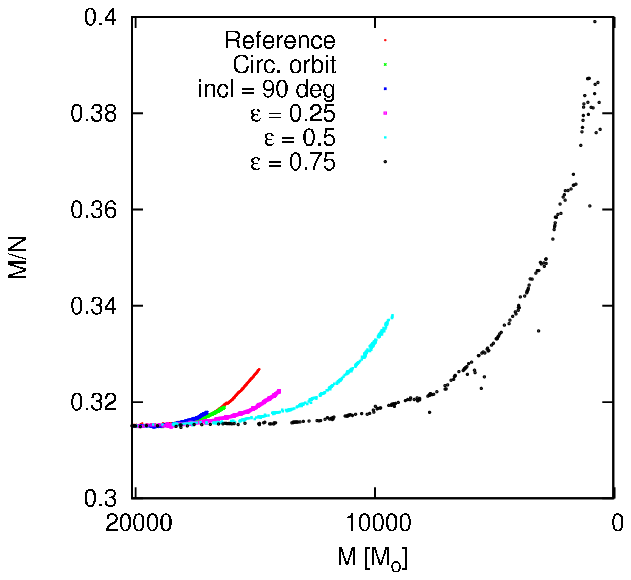}
  \caption{The average mass of stars, $M/N$, versus bound mass, $M$, for the clusters at an initial galactocentric distance of 4.25 kpc (left), 12.75 kpc (middle) and 17 kpc (right). The preferential loss of low mass stars is very pronounced in most clusters. Since preferential loss is a phenomenon of two-body relaxation, the earlier the increase in $M/N$ with decreasing $M$ the more important is two-body relaxation in the cluster.}
  \label{MNrgal}
\end{figure*}
\begin{figure*}
\includegraphics[width=168mm]{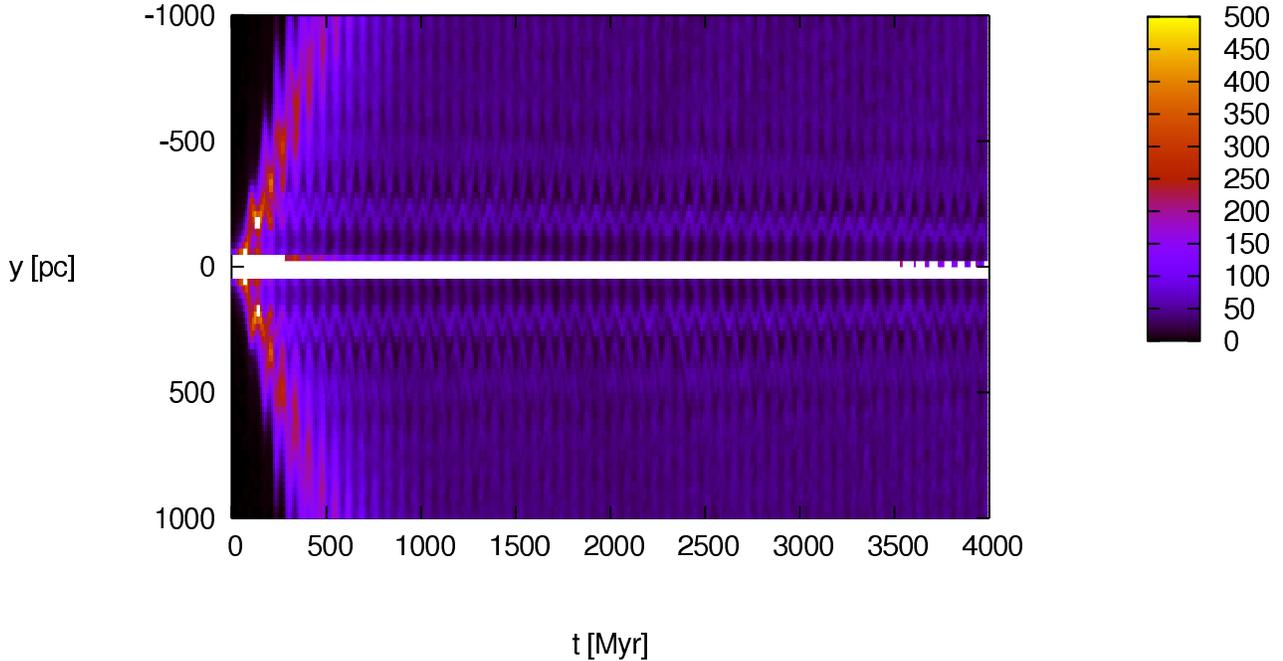}
  \caption{As Fig.~\ref{yc_ref}: time evolution of the number distribution of stars along the tidal tails in bins of 25 pc for the cluster with apogalactic distance of 4.25 kpc and an eccentricity of 0.25 (discussed in Sec. \ref{ssec:fgprgal}). Three to four orders of overdensities can be seen in this diagram, periodically changing in distance due to the eccentricity. The clusters at this apogalactic distance have the smallest radii and therefore the shortest relaxation times. The periodic influence of the tidal variation does not have a large effect on the cluster, such that the escape conditions for the formation of epicyclic overdensities are still well fulfilled for a majority of escapers.}
  \label{yc_4250}
\end{figure*}
\begin{figure}
\includegraphics[width=84mm]{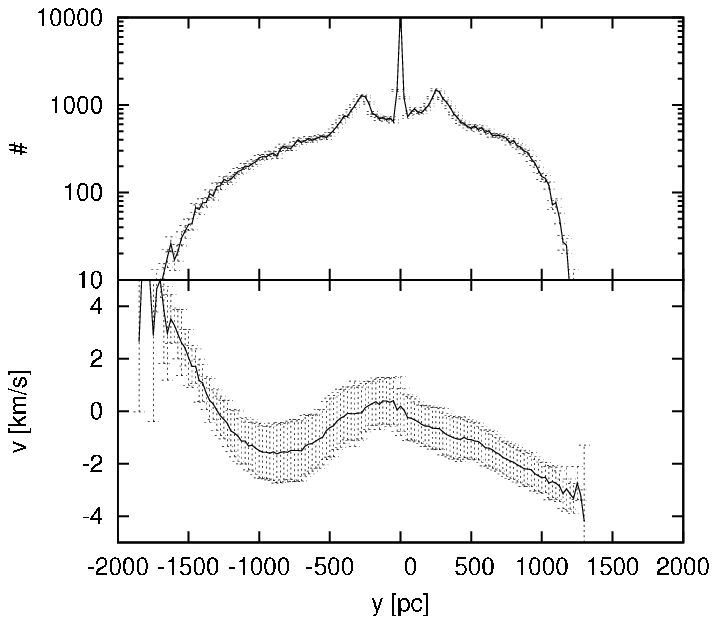}
  \caption{Number distribution of stars along the tidal tails in bins of 25 pc (upper panel) and corresponding mean velocities of the stars within the bins (with respect to the orbital velocity of the cluster, eq.~\ref{eq:v}) for the cluster with the most extreme tidal conditions in our set, i.e. an apogalactic distance of 4.25 kpc and an eccentricity of 0.75 (discussed in Sec. \ref{ssec:fgprgal}). The snapshot was taken at $t=140$ Myr when the cluster had a mass of about $3000 \msun$. Error bars in the upper panel are Poisson errors and in the lower panel show the velocity dispersion within the corresponding bin (eq.~\ref{eq:sigma}).  Notice the different scale on the y-axis compared to the other figures, which is necessary as mass loss from this cluster is more intense. But even though escape from this cluster is happening with a wide range of escape conditions epicyclic overdensities are still visible in the tails.}
  \label{y_4250}
\end{figure}
To see whether the behaviour of the cluster, and especially of the tails, changes when the strength and the duration of the pericentre passages and disk shocks are varied, we investigate all of the above models at three further galactocentric distances of 4.25, 12.75 and 17 kpc. The initial conditions of the test cluster are adjusted to the given tidal field, i.e. a Plummer model with a half-mass radius such that $R_h/R_{tide}=0.2$, since \citet{Tanikawa05} as well as \citet{Gieles08} have shown that the ratio of half-mass and tidal radius significantly determines the evolution of a cluster. Hence, our setup is a reasonable choice for a good comparability. On the other hand, due to the fixed ratio of half-mass radius and tidal radius we simultaneously change the two-body relaxation time-scale with these adjustments.

Most globular clusters of the Milky Way lie below the ratios studied here \citep{Harris96}, i.e. are more compact and, hence can resist tidal perturbations more easily and have a smaller two-body relaxation time.

\subsubsection{Internal evolution}
In Fig.~\ref{Mrgal} the mass evolution of all clusters at 4.25, 12.75 and 17 kpc is shown. As mentioned above, since all clusters have the same initial concentration the two-body relaxation time decreases for the clusters closer to the galactic centre. Therefore, the dissolution of the models on circular orbits is stronger the closer the model is to the galactic centre.

Moreover, an increasing influence of pericentre passages for decreasing galactocentric radii can be seen when comparing the models on eccentric orbits at a given apogalactic distance with the respective circular model. Models with a smaller apogalactic distance go deeper into the bulge at perigalacticon and hence are more strongly influenced by the tidal-field variations as the angular velocity and hence the stress on the cluster grows with decreasing galactocentric radius.

In Fig.~\ref{MNrgal} the mean stellar mass within the clusters is plotted versus the bound mass left in the cluster. An earlier increase in mean mass shows that the two-body relaxation time scale is short compared to the time scale of dissolution of the cluster. Therefore, the probability of observing epicyclic overdensities grows for models with an earlier increase in mean mass. As we can see from the figure, all models show an increase in mean stellar mass. Especially at 4.25 kpc, where the two-body relaxation time is shortest, most models show a steep increase in mean stellar mass and epicyclic overdensities are therefore expected.

From our sample just the model at 4.25 kpc with eccentricity of 0.75 dissolves so quickly (Fig.~\ref{Mrgal}) and shows so little increase in the mean mass (Fig.~\ref{MNrgal}) that we would not expect observing epicyclic overdensities in this case.

\subsubsection{Evolution of the tidal tails}

As shown in KMH, in the case of a cluster in a constant tidal field, i.e. on a circular orbit, epicyclic overdensities are generated by stars which escape from the cluster as a result of two-body relaxation. The tidal variations studied in this work enhance the mass loss of the investigated clusters and also the scatter in escape conditions. For most orbital parameters two-body relaxation seems to be actively involved in the dissolution of the cluster such that the influence of the tidal variations can not be too violent.

We indeed find epicyclic overdensities for all our models. In Fig.~\ref{yc_4250} we show as an example the disk-shock model at 4.25 kpc in a time series of the number density along the tidal tails. The figure shows two pronounced orders of overdensities which oscillate, as the cluster is accelerated and decelerated by the galactic disk, and in mean decrease in distance as the cluster loses mass.

Contrary to our expectations, even the most extreme cluster in our sample, the one at 4.25 kpc with an eccentricity of 0.75, is able to grow epicyclic overdensities. Its life time is very short; the cluster survives for only about three revolutions. Still this is sufficient for most of the escaping stars to go through at least one epicyclic loop.

In Fig.~\ref{y_4250} this cluster is shown at $t=140$ Myr when it has a mass of about 3000 $\msun$ and has completed about 5/4 revolutions about the galaxy (one epicycle of an escaping star takes about 1/2 revolution). The mean stellar density in the tails in this case is comparatively high as many stars are lost from the cluster, such that we had to increase the scale on the y-axis. The first-order epicyclic overdensities are prominently imprinted. The orbital velocity of stars shows no clear variation along the tails except the large-scale variation due to the orbit about the galaxy. The velocity dispersion in the corresponding bins are quite large, though. This indicates that the scatter in escape conditions is large. Nevertheless, there must be a dominant speed and a dominant radius from which stars escape such that they interfere constructively and the observed overdensities can build up.

In the end, our numerical investigation has convincingly shown that we have to anticipate epicyclic overdensities whenever we observe tidal tails.

\section{Conclusions}
Our numerical investigation has shown that star clusters always grow epicyclic overdensities of the kind predicted in KMH. We were able to prove this for a large variety of orbits and for a test cluster which is comparatively fluffy (large ratio of half-mass radius to tidal radius) and hence more vulnerable to tidal influences than most star clusters of the Milky Way. Only in very extreme conditions, if the cluster gets literally disrupted by a single shock, the overdensities would not be able to arise, as they need a certain time (about one to two orbits) to build up. Since this would imply that such a cluster has a very low chance of being observed, we can conclude that epicyclic overdensities can be found in any star-cluster tidal tail on the sky.

In KMH we argued that the epicyclic overdensities arise if a majority of stars leaves the cluster with evaporative escape conditions, i.e. with a velocity just slightly above the escape velocity when escaping from the cluster through one of the two Lagrange points. For clusters on circular orbits these conditions are established via two-body relaxation. Pericentre passages and disk shocks perturb the evolution of a cluster and increase the scatter in escape conditions. Nevertheless, we observe epicyclic overdensities for all tested clusters. The influence of pericentre passages and disk shocks on the escape conditions therefore has to be rather moderate for the investigated range of orbital parameters. Since most clusters of the Milky Way are more compact than our test cluster, the average influence of eccentric orbits or disk shocks on the escape conditions should be even weaker than in our investigation.

In KMH we showed that the distance from the cluster centre to the first epicyclic overdensity, $y_C$, is a multiple of the tidal radius. Hence, as the cluster constantly loses mass and the tidal radius consequently gets smaller, $y_C$ monotonically decreases with time. Here we showed that this is just the case for circular orbits, since for eccentric orbits the periodic acceleration and deceleration of the tidal tails leads to a stretching and compression of the whole system and therefore periodically increases and decreases $y_C$ (e.g. Fig.~\ref{yc_e025}). At apocentre $y_C$ is minimal as the system is maximally compressed while at pericentre $y_C$ reaches its maximum value. The strength of the variation in $y_C$ depends on the eccentricity and is a factor of about 7.5 for an eccentricity of 0.75 (Fig.~\ref{yct_ecc}). However, on the mean, $y_C$ decreases with time as the cluster evaporates.

We furthermore provide an approximation for estimating the distance $y_C$ in the case of time-variable tidal fields (Sec.~\ref{ssec:yc of t}). It is based on the assumption that the cluster experiences a mean tidal field on its orbit, as it is not able to react to the tidal variations in time. This assumption also suggests that escape from a cluster in a time-dependent tidal field happens from the ``edge'' of the cluster, which is comparable to the apogalactic tidal radius, rather than from a theoretically determined tidal radius. K\"upper, Kroupa, Baumgardt \& Heggie (in preparation) show that the radius of the ``edge'' of a cluster in the given range of orbital parameters stays approximately constant during one period. Hence, the assumptions are justified and the theoretical predictions agree well with the $N$-body results.

Moreover, we found further over- and underdensities which arise through the non-uniform acceleration of the stars within very extended tidal tails; since tails may easily span several kpc, different parts may experience very different forces and have significantly different velocities. This leads to a non-uniform stretching and compression of the tidal tails causing a large-scale variation of the stellar density along the tails (Fig.~\ref{yc_e05}). This effect may be very helpful for determining the orbit of an observed cluster like Pal~5.

Throughout our investigation we were not able to detect overdensities which could be attributed to a pericentre passage or a disk shock, i.e. a bunch of stars which gets unbound during one tidal event and is then moving in a group along the tail. Furthermore, we have shown that the mass-loss rate is approximately constant for all kinds of orbits when the tidal radius is calculated using eq.~\ref{eq:rtide}. The mass-loss rate is larger for more eccentric orbits or orbits with disk shocks closer to the galactic centre, but does not cause steps in the mass curve, if the pericentre dips are neglected (Fig.~\ref{M8500}). This is due to a delay in escape of stars which get energetically unbound through a tidal event. We argue that the stars which are most affect by these events orbit the cluster at radii comparable to the apogalactic tidal radius of the cluster. The orbital times of those stars within the cluster is of the same order as the orbital time of the cluster within the galaxy. Escape therefore does not happen instantaneously, but the stars leave the cluster delayed with respect to the tidal event, where the delay of a specific star may last several orbital times (Fig.~\ref{1224KR}). As a consequence of this, we furthermore suggest to use the apogalactic tidal radius for the computation of internal quantities rather than the perigalactic tidal radius as has often been done in numerical studies.

The Milky-Way globular cluster Palomar~5 is the only cluster with prominent, extended tidal tails which, furthermore, show significant substructure. The most detailed numerical investigation of this cluster has been performed by \citet{Dehnen04} but they were not able to reproduce the substructure within their models. Instead of adding substructure to the Milky-Way potential like spiral arms or giant molecular clouds to increase the time variations of the tidal field, or adding dark-matter sub-haloes to the Milky-Way potential to perturb the dynamically cold tidal stream, as suggested by Dehnen et al., we suggest that most overdensities within the tails of Pal~5 are of epicyclic origin or are due to an asymmetric acceleration along the tails.

Another interesting object for this kind of analysis is the recently discovered GD-1 stream\citep{Grillmair06b, Koposov09}, which is thin and thus kinematically cold. Even though the progenitor of this stream has not been found yet it is most probably emanating from a globular cluster which may have dissolved by now. This stream also shows substructure which may help to locate the position of the cluster within the stream since the lowest orders of epicyclic overdensities (the overdensities closest to the cluster) are always most prominent.

There have also been stellar overdensities discovered without any trace of a tidal stream or an object to which they can be attributed to, e.g. the Bo\"otes II dwarf spheroidal \citep{Koch09}. With ongoing and future surveys looking for stellar overdensities in the Milky-Way halo, the number of such objects may well increase. To ease the identification of these objects as epicyclic overdensities within tidal tails which are themselves below the detection limit, we are going to provide a detailed investigation with predictions for observers including such factors as radial velocity signatures, the stellar mass function, typical sizes and masses, etc., in a future contribution.

\section*{Acknowledgements}
AHWK would like to thank Ulf L\"ockmann for mathematical support. Furthermore, the authors would like to thank Sverre Aarseth for making his \textsc{NBODY} codes freely available and an anonymous referee for useful suggestions. This research was supported in part by the National Science Foundation under Grant No. PHY05-51164.

\bsp

\label{lastpage}
\end{document}